\documentclass{emulateapj}

\usepackage{ulem}

\slugcomment{Published in \apj,  2013 May 29}

\shorttitle{}
\shortauthors{}

\begin{document}

\title{Optical-to-Near-Infrared Simultaneous Observations for the Hot Uranus GJ3470b: A Hint for Cloud-free Atmosphere}

\author{Akihiko Fukui\altaffilmark{1}, Norio Narita\altaffilmark{2}, Kenji Kurosaki\altaffilmark{3}, Masahiro Ikoma\altaffilmark{3},  Kenshi Yanagisawa\altaffilmark{1}, Daisuke Kuroda\altaffilmark{1}, Yasuhiro Shimizu\altaffilmark{1}, Yasuhiro H. Takahashi\altaffilmark{2,4}, Hiroshi Ohnuki\altaffilmark{5}, Masahiro Onitsuka\altaffilmark{6}, Teruyuki Hirano\altaffilmark{7}, Takuya Suenaga\altaffilmark{6}, Kiyoe Kawauchi\altaffilmark{2}, Shogo Nagayama\altaffilmark{2}, Kouji Ohta\altaffilmark{8}, Michitoshi Yoshida\altaffilmark{9}, Nobuyuki Kawai\altaffilmark{10}, Hideyuki Izumiura\altaffilmark{1}}

\altaffiltext{1}{Okayama Astrophysical Observatory, National Astronomical Observatory of Japan, 
Asakuchi, Okayama 719-0232, Japan}
\email{afukui@oao.nao.ac.jp}
\altaffiltext{2}{National Astronomical Observatory of Japan, 2-21-1 Osawa, Mitaka, Tokyo 181-8588}
\altaffiltext{3}{Department of Earth and Planetary Science, The University of Tokyo, 7-3-1 Bunkyo-ku, Tokyo 113-0033, Japan}
\altaffiltext{4}{Department of Astronomy, The University of Tokyo, 7-3-1 Hongo, Bunkyo-ku, Tokyo 113-0033}
\altaffiltext{5}{Department of Earth and Planetary Sciences, Tokyo Institute of Technology, 2-12-1 Ookayama, Meguro-ku, Tokyo 152-8551}
\altaffiltext{6}{The Graduate University for Advanced Studies, 2-21-1 Osawa, Mitaka, Tokyo 181-8588}
\altaffiltext{7}{Department of Physics, The University of Tokyo, 7-3-1 Hongo, Bunkyo-ku, Tokyo 113-0033}
\altaffiltext{8}{Dept. of Astronomy, Kyoto University, Kitashirakawa-Oiwake, Sakyo, Kyoto, 606-8502, Japan}
\altaffiltext{9}{Hiroshima Astrophysical Science Center, Hiroshima University 1-3-1, Kagamiyama, Higashi-Hiroshima, Hiroshima, 739-8526, Japan}
\altaffiltext{10}{Dept. of Physics, Tokyo Institute of Technology, 2-12-1, Oookayama, Meguro, Tokyo, 152-8551, Japan}

\begin{abstract}
We present optical ($g'$, $R_\mathrm{c}$, and $I_\mathrm{c}$) to near-infrared ($J$) simultaneous photometric observations for a primary transit of GJ3470b, a Uranus-mass transiting planet around a nearby M dwarf, by using the 50-cm MITSuME telescope and the 188-cm telescope both at Okayama Astrophysical Observatory. From these data, we derive the planetary mass, radius, and density as $14.1 \pm 1.3$ $M_\oplus$, $4.32^{+0.21}_{-0.10}$ $R_\oplus$, and  $0.94 \pm 0.12$ g~cm$^{-3}$, respectively, thus confirming the low density that was
reported by Demory et al.  based on the $Spitzer$/IRAC 4.5-$\mu$m photometry ($0.72^{+0.13}_{-0.12}$ g~cm$^{-3}$). 
Although the planetary radius is about 10\% smaller than that reported by Demory et al., 
this difference does not alter their conclusion that the planet possesses a hydrogen-rich envelope whose  mass is approximately 10\% of the planetary total mass. 
On the other hand, we find that the planet-to-star radius ratio ($R_p/R_s$) in the $J$ band ($0.07577^{+0.00072}_{-0.00075}$) is smaller than that in the $I_\mathrm{c}$ ($0.0802 \pm 0.0013$) and 4.5-$\mu$m ($0.07806^{+0.00052}_{-0.00054}$) bands by $5.8\% \pm 2.0$\%  and $2.9\% \pm 1.1$\%, respectively.
A plausible explanation for the differences is that the planetary atmospheric opacity varies with wavelength
due to absorption and/or scattering by atmospheric molecules.
Although the significance of the observed $R_p/R_s$ variations is low, if confirmed, this fact would suggest that GJ3470b does not have a thick cloud layer in the atmosphere.
This property would offer a wealth of opportunity for future transmission-spectroscopic observations of this planet to search for certain molecular features, such as H$_2$O, CH$_4$, and CO, without being prevented by clouds.

\end{abstract}


\section{INTRODUCTION}
Transiting extrasolar planets provide not only their masses and radii but also other valuable information on the planets such as atmospheric constituents.
Because the optical thickness of a planetary atmosphere varies with wavelength depending on the atmospheric composition, one can probe atmospheric constituents by measuring transit radii with different wavelengths \citep[e.g.][]{2000ApJ...537..916S,2001ApJ...553.1006B}. 

So far, this technique, known as the transmission spectroscopy, has been applied for several transiting hot Jupiters orbiting nearby bright host stars, e.g., HD~209458b and HD~189733b.
In the atmosphere of HD~209458b, past observations detected many absorption and scattering features such as Na \citep{2002ApJ...568..377C,2008A&A...487..357S}, H \citep{2003Natur.422..143V}, O, C \citep{2004ApJ...604L..69V}, TiO, VO \citep{2008A&A...492..585D}, H$_2$ \citep{2008A&A...485..865L}, H$_2$O \citep{2007ApJ...661L.191B,2010MNRAS.409..963B}, and CO \citep{2010Natur.465.1049S}. On the contrary, HD~189733b has been revealed to have a featureless transmission spectrum over optical to possibly near infrared regions \citep{2008MNRAS.385..109P,2009A&A...505..891S,2011A&A...526A..12D,2012MNRAS.422..753G}, while absorption features of Na and H were detected \citep{2008ApJ...673L..87R,2010A&A...514A..72L}. This featureless spectrum has been interpreted as Rayleigh scattering due to  high-altitude haze which dominates over molecular features \citep{2008A&A...481L..83L}.  

Recently, it has become possible to expand this technique to low-mass planets, often referred to as exo-Neptunes (10 $\lesssim$ $M_p/$M$_\oplus$ $\lesssim$ 30) and super-Earths ($M_p/$M$_\oplus$ $\lesssim$ 10), thanks to the discoveries of transiting low-mass planets around nearby low-mass stars (M dwarfs). Because  M dwarfs have smaller radii compared to Sun-like stars, they show deeper transits for a same-sized transiting planet. This enables us to obtain high signal-to-noise-ratio transit signals even for small planets.
The first two such examples are GJ436b \citep{2004ApJ...617..580B, 2007A&A...472L..13G} and GJ1214b \citep{2009Natur.462..891C}. They can be thought to be the representatives of exo-Neptunes and super-Earths, respectively, due to their masses ($\sim$23 and $\sim$6.6 $M_\oplus$, respectively).
As for GJ436b, no firm molecular feature has been detected by transmission spectroscopy probably because of difficulties due to 
instrumental systematics \citep{2011MNRAS.411.2199G} and stellar activity \citep{2011ApJ...735...27K}. On one hand,  a methane-poor and CO-rich atmosphere in chemical disequilibrium is suggested based on the emission spectrum obtained via secondary-eclipse observations \citep{2010Natur.464.1161S}, while its chemical condition is  debated by \citet{2011ApJ...731...16B} and \citet{2011ApJ...738...32L}.
The super-Earth GJ1214b has received much more attention since its discovery. Intensive observations by many observational groups have revealed that GJ1214b has a flat spectrum over optical to infrared wavelengths 
\citep[e.g.][]{2011ApJ...731L..40D,2011ApJ...743...92B,2012ApJ...747...35B,2012A&A...538A..46D,2012arXiv1210.3169N,2013ApJ...765..127F}. 
This flat spectrum has been interpreted as the consequence of either a water-dominated atmosphere or a hydrogen-dominated but hazy/cloudy atmosphere, although which model is correct is still an open question \citep[e.g.][]{2012ApJ...756..176H}.

GJ3470b, the target of this paper, is the third low-mass ($< 30 M_\oplus$) transiting planet discovered among nearby ($< 35$ pc) M dwarfs \citep{2012A&A...546A..27B} which provides a great opportunity to extend the atmospheric study of low-mass planets.
Because its mass, $\sim14 M_\oplus$, is intermediate between those of GJ436b and GJ1214b, this planet should be useful for a comparative study of the atmospheric properties of low-mass planets. 
\citet[][hereafter D13]{2013arXiv1301.6555D} have recently reported follow-up observations for this system including $Spitzer$/IRAC 4.5-$\mu$m photometry of two primary transits of GJ3470b.
They precisely determined the planetary mass and radius as   $13.9^{+1.5}_{-1.4} M_\oplus$ and $4.83^{+0.22}_{-0.21} R_\oplus$, respectively, revealing  its very low density ($\rho_p = 0.72^{+0.13}_{-0.12}$ g cm$^{-3}$). 
This implies that  GJ3470b has a light-gas atmosphere with enlarged spectrum features,  thus making this planet an attractive target for studies of transmission spectroscopy.  

Here we present optical-to-near-infrared simultaneous photometric observations for a primary transit of GJ3470b obtained by using the 188-cm telescope and the 50-cm MITSuME telescope, both at the Okayama Astrophysical Observatory.
The simultaneous observations have a great advantage for transmission-spectroscopic study. If the host star has cool starspots on its surface, then the transit depth (square of planet-star radius ratio) can vary with time due to the variations of the apparent luminous area of the star according to the stellar rotation and appearing/vanishing of the starspots \citep[e.g.][]{2008MNRAS.385..109P}. This changes the apparent planetary radius with time. Therefore, the simultaneous observations enable us to investigate the wavelength dependence of the planetary radius without concern for this starspot effect.

The rest of this paper is organized as follows. We describe our observations in Section \ref{sec:obs}, followed by data reduction and analysis
 in Section \ref{sec:data_reduction} and Section \ref{sec:mcmc}, respectively. We  discuss the implications of  our results in Section \ref{sec:discussion} and summarize this study in Section \ref{sec:summary}.

\section{OBSERVATIONS}
\label{sec:obs}

\subsection{ISLE $J$-band Observations}
\label{sec:isle_obs}
We conducted $J$-band photometric observations for GJ3470
on the expected transit night of 2012 November 15
by using the near-infrared imaging and spectroscopic instrument ISLE \citep{2006SPIE.6269E.118Y,2008SPIE.7014E.106Y}, which is mounted on the Cassegrain focus of the 188 cm telescope at the Okayama Astrophysical Observatory in Japan.
ISLE has a 1024 $\times$ 1024 HgCdTe HAWAII-1 array with a pixel scale of 0$''$.245 pixel$^{-1}$, providing a field of view (FOV) of 4$'$.3 on a side.
In order to perform relative photometry, we introduced a comparison star (TYC 1363-2087-1) onto the detector simultaneously with the target star GJ3470; the two stars are separated by 2$'$.9. The comparison star has a similar brightness in the $J$ band ($J$=8.73) to  that of GJ3470 ($J$=8.79), while it has a different color ($V-J$=1.60) to that of GJ3470 ($V-J$=3.54). Although this color difference causes a systematic trend in the relative-photometric light curve due to the differential-color extinction effect, this trend can properly be corrected, as discussed in Section \ref{sec:reduction_isle}.
The photometric properties of the target and comparison stars are listed in Table \ref{tbl:list_compstars}.  We note that we used only the one comparison star because there is no other star that has a similar brightness to GJ3470 within two magnitudes in the $J$ band in the FOV.
We also note that this comparison star shows no short-term variability, as shown in Section \ref{sec:reduction_isle}.
At the beginning of the observations, the stellar positions on the detector were carefully set so that the stellar images did not cover any bad pixels. 
During the observations, the telescope was defocused so that the full width at half maximum of the stellar point spread function (PSF) was 23--30 pixels, or 5$''$.6--7$''$.4, in order to compensate for an imperfect flat-fielding correction for pixel-to-pixel-sensitivity variations and to extend the exposure time as long as possible while avoiding saturation \citep[see e.g.][]{2009MNRAS.396.1023S}.
The exposure time was set to 30 s, with which the peak analog-to-digital-unit (ADU) count of the brighter star (the comparison star) was $\sim$20,000 at most. This is well below the threshold of 25,000 ADU, above which the AD converting relation deviates more than 1\% from a linear relation. The dead time including readout time for each exposure  was 6 s. 
The observations spanned 4.4 hr, covering the entire transit (1.9 hr) as well as 1.8 hr prior to and 0.75 hr posterior to the transit.
The weather was clear, except for about 50 minutes in the course of the pre-transit time when clouds had passed.
An observing log is shown in Table \ref{tbl:obslog}.

\begin{deluxetable*}{lccccc}
\tabletypesize{\footnotesize}
\tablecaption{Photometric properties of the target and selected comparison stars \label{tbl:list_compstars}}
\tablewidth{15cm}
\tablehead{
\colhead{} & \colhead{Name} & \colhead{$g ^{(1)}$} &  \colhead{$r ^{(1)}$} & \colhead{$i ^{(1)}$} & \colhead{$J ^{(2)}$}
}
\startdata
Target star & GJ3470 & 15.44 & 11.87 & 10.76 & 8.79\\[3pt]
\hline\\[-5pt]
Comparison star  & TYC 1363-2087-1 & 11.34 & 10.41 &  10.07 & 8.73\\[3pt]
for $J$ band & &  & & &\\[3pt]
\hline\\[-5pt]
 & BD+15 1718 & 10.33 & 9.85 & 9.75 & 8.92\\[3pt]
Comparison stars & TYC 1363-2087-1 & 11.34 & 10.41 &  10.07 & 8.73\\[3pt]
 for $I_\mathrm{c}$ band & TYC 1363-1635-1 & 11.53 & 10.71 & 10.41 & 9.16\\[3pt]
 & TYC 1363-1897-1 & 12.03 & 11.08 & 10.82 & 9.53\\[3pt]
\hline \\[-5pt]
 & TYC 1363-1897-1 & 12.03 & 11.08 & 10.82 & 9.53\\[3pt]
  & TYC 1363-2033-1 & 14.73 & 11.79 & 11.60 & 11.23\\[3pt]
 & SDSS9 J075830.49+152215.9 &  12.94 & 11.91 & 11.60 & 10.26\\[3pt]
  & SDSS9 J075912.94+151859.8 & 13.74 & 11.94 & 11.85 &  11.64\\[3pt]
Comparison stars & SDSS9 J075910.70+151910.3 & 12.19 & 12.52 & 14.76 &  11.62\\[3pt]
for $R_\mathrm{c}$ band & SDSS9 J075850.52+152743.7 & 13.70 & 12.55 & 12.53 &  11.40\\[3pt]
 & SDSS9 J075918.34+152709.3 & 13.18 & 12.84 & 12.72 &  11.81\\[3pt]
   & SDSS9 J075926.99+152250.4 & 14.26 & 14.03 & 13.77 &  10.51\\[3pt]
 & SDSS9 J075854.12+153057.1 & 15.43 & 14.38 & 11.63 &  10.04\\[3pt]
 & SDSS9 J075901.08+151640.7 & 14.88 & 14.77 & 14.25 &  10.87\\[3pt]
\hline \\[-5pt]
 & BD+15 1718 & 10.33 & 9.85 & 9.75 & 8.92\\[3pt]
& TYC 1363-2087-1 & 11.34 & 10.41 &  10.07 & 8.73\\[3pt]
 & TYC 1363-1635-1 & 11.53 & 10.71 & 10.41 & 9.16\\[3pt]
Comparison stars & TYC 1363-2233-1&12.06 & 10.56 & 9.03 &  6.65\\[3pt]
 for $g'$ band   & SDSS9 J075923.58+153016.5 & 13.30 & 12.83 & 12.75 & 11.81\\[3pt]
 & SDSS9 J075850.52+152743.7 & 13.70 & 12.55 & 12.53 &  11.40\\[3pt]
  & SDSS9 J075926.99+152250.4 &  14.26 & 14.03 & 13.77 &  10.51\\[3pt]
 & SDSS9 J075901.08+151640.7 & 14.88 & 14.77 & 14.25 &  10.87
\enddata
\tablenotetext{}{{\bf References.} (1) SDSS DR9 \citep{2012ApJS..203...21A}; (2) Two Micron All Sky Survey \citep{2003yCat.2246....0C}}
\end{deluxetable*}

\begin{deluxetable*}{lccccc}
\tablewidth{15cm}
\tablecaption{Observing Log\label{tbl:obslog}}
\tablehead{
\colhead{Observing Date} & \colhead{Filter} & \colhead{Telescope} & \colhead{Exp. Time} & \colhead{$N_\mathrm{all}$}\tablenotemark{a} & \colhead{Air Mass Variation}\\
& & & (s) & &
}
\startdata
2012 November 15 & $J$ & OAO 188 cm & 30 & 352 & 1.42 $\rightarrow$ 1.06 $\rightarrow$ 1.12\\
2012 November 15 & $I_\mathrm{c}$ & MITSuME 50 cm & 60 & 165 & 1.28 $\rightarrow$ 1.06 $\rightarrow$ 1.11\\
2012 November 15 & $R_\mathrm{c}$ & MITSuME 50 cm & 60 & 162 & 1.28 $\rightarrow$ 1.06 $\rightarrow$ 1.11\\
2012 November 15 & $g'$ & MITSuME 50 cm & 60 & 164 & 1.28 $\rightarrow$ 1.06 $\rightarrow$ 1.11
\enddata
\tablenotetext{a}{\ The total observed data points used for analyses, i.e., after omitting outliers.}
\tablenotetext{}{}
\end{deluxetable*}

\begin{figure}[h]
\begin{center}
\includegraphics[width=8.5cm]{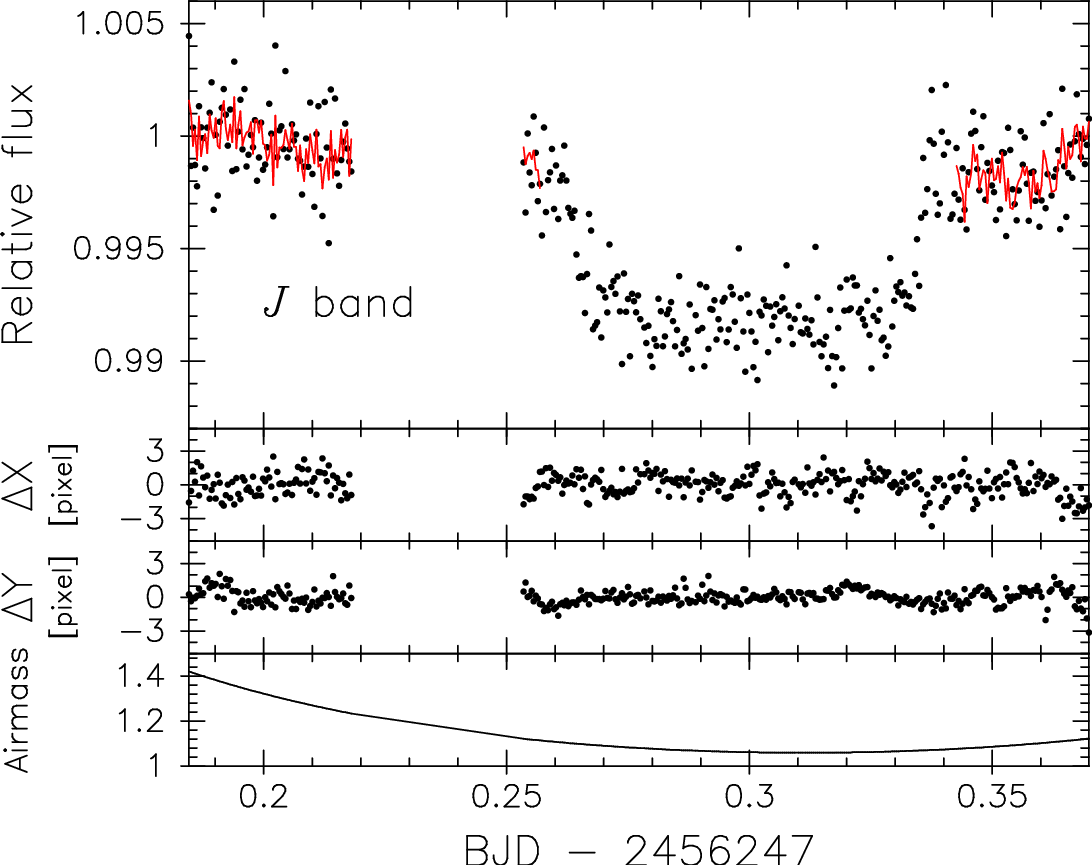}\\
\caption{Top panel: the $J$-band uncorrected light curve of GJ3470b. The best-fit baseline model, which is as a function of air mass and stellar centroid displacements in X and Y directions, is over plotted as solid line in the out of transit (OOT) part. Second panel: the stellar centroid displacements in X direction. Third panel: the same as the second panel but in Y direction. Bottom panel: the air mass variations.   \label{fig:rawlc_isle}}
\end{center}
\end{figure}

The 188-cm telescope is equipped with an offset guider system on the Cassegrain focus.
However,  it had been recognized from past observations with ISLE that the stellar centroid positions on the ISLE detector slightly (a few pixels) changed over several hours even when the guiding system was activated\footnote{We attributes the cause of this stellar displacement to a relative mechanical offset between the ISLE detector and the guiding CCD camera depending on the telescope position, rather than other effects such as the differential atmospheric refraction between infrared and optical wavelengths.}. 
Generally, a stellar displacement on a detector introduces some amount of systematic errors into photometry \citep[e.g.][]{2005ApJ...626..523C}.
In order to avoid this effect, we have developed an offset-correcting system which corrects the reference point of the guide star on the guiding camera, by calculating the displacement of stellar centroids on the last ISLE images relative to a reference image. Each correction needs an additional dead time for ISLE of 8 s. 
For the transit observations of GJ3470b, 
we corrected the offset as often as once in 10 ISLE exposures.
As a result, the stellar centroid change during the observations was well suppressed, with 1.1 and 0.6 pixels in root mean square (rms) for the X (right ascension) and Y (declination) directions, respectively.
The centroid variation with time on the night is shown in Figure \ref{fig:rawlc_isle}.

\subsection{MITSuME Optical Observations}

Simultaneously with the ISLE $J$-band observations, we also conducted photometric observations for the transit of GJ3470b in optical wavelengths by using the 50-cm MITSuME telescope at the Okayama Astrophysical Observatory.
The telescope is equipped with three 1024 $\times$ 1024 pixels CCD cameras, enabling it to obtain $I_\mathrm{c}$-, $R_\mathrm{c}$-, and $g'$-band images simultaneously \citep{2005NCimC..28..755K,2010AIPC.1279..466Y}. Each camera has a pixel scale of 1$''$.5 pixel$^{-1}$ for a FOV of 26$' \times$ 26$'$. 

At the beginning of the observations, we set the FOV so that the target star was nearly the center of the respective CCDs. Thanks to the wide FOV, there were dozens of stars brighter than 15 mag in each band in the FOV that could potentially be good comparison stars for relative photometry.
The telescope was defocused so that the FWHM of the stellar PSF is about 10 pixels, or $\sim$15$''$, for the same purpose as the ISLE observations. We note that the contamination light from objects surrounding the target star is negligible because there is no object brighter than 20 and 16 magnitudes  in any band within 10$''$ and 20$''$, respectively, from GJ3470 in the Sloan Digital Sky Survey (SDSS) photometric catalog \citep{2011yCat.2306....0A}.
We started the MITSuME observations about 35 minutes behind the ISLE observations.
The exposure time was set to 60 s. The dead time including readout time for each exposure was 3 s for all bands. The observing log is compiled in Table \ref{tbl:obslog}.

Because this telescope has no mechanical auto-guiding system and previously caused a large tracking error ($\sim$100 pixels) over several hours, 
we have developed a self-guiding software that calculates the displacement of the stellar centroid positions on the last-observed $I_\mathrm{c}$-band image relative to a reference image, then feeds it back to the telescope to correct the tracking error soon after the last image is obtained (within a few seconds).
By using this self-guiding software, the stellar centroid displacement during the observations for GJ3470 were kept with 1.4 and 0.5 pixels in rms for the X (right ascension) and Y (declination) directions, respectively, for all the three bands.

\section{DATA REDUCTION}
\label{sec:data_reduction}

\subsection{Reduction and Baseline Correction for the ISLE data}
\label{sec:reduction_isle}

The obtained ISLE images are reduced with the standard procedure, including dark-image subtraction and flat-fielding correction.
For the flat-fielding correction, 100 dome-flat images obtained on the observing night are used to create a flat-fielding image.
Then, aperture photometry is performed for the target and comparison stars on the reduced images  by using a customized tool \citep{2011PASJ...63..287F}, applying a constant aperture radius for both the target and comparison stars for all images.
A light curve is created by dividing GJ3470's fluxes by those of the comparison.
We eliminate apparent outliers due to such factors as passing clouds and cosmic-ray hitting from the light curve by checking the reduced images carefully.
The time for each data point is assigned as the mid time of each exposure in the time system of
Barycentric Julian Day (BJD) based on Barycentric Dynamical Time (TDB), which is converted from the time stamp recorded on the FITS header in the time system of Julian Day (JD) based on Coordinated Universal Time (UTC) by using the code of \citet{2010PASP..122..935E}.
We remind the readers that the time standard of UTC should not be used for time-critical studies such as transit timing variations (TTVs), because it is discontinuous due to interruptions by occasional reap second. Alternatively, using TDB is recommended by \citet{2010PASP..122..935E}, as it is continuous and precise at $<$3.4 ms level (ultimately $\sim$1 $\mu$s level if higher order effects are corrected).

\begin{deluxetable*}{lccccccc}
\tablewidth{17cm}
\tablecaption{Fitting results for $J$-band data for different baseline models\label{tbl:amoeba} \tablenotemark{a}}
\tablehead{
\colhead{Variables} & \colhead{BIC$_\mathrm{oot}$} & \colhead{rms$_\mathrm{oot}$} & \colhead{$\chi^2_\mathrm{all}$ \tablenotemark{b}} & \colhead{$R_p/R_s$ \tablenotemark{c}} & \colhead{$a/R_s$ \tablenotemark{c}} & \colhead{$b$ \tablenotemark{c, d}} & \colhead{$T_\mathrm{c}$ \tablenotemark{c}}\\
\colhead{$\{{\bf X}\}$} & & ($\times 10^{-3}$) & & ($\times 10^{-2}$) & & & \colhead{[BJD$_\mathrm{TDB}$-2456247]}
}
\startdata
$z$ & 224.2  & 1.61 & 449.3 & 7.754 $\pm$ 0.098 & 14.91 $^{+1.2}_{-0.33}$ & 0.00 $\pm$ 0.21 & 0.29949 $\pm$ 0.00025\\[3pt]
$\mbox{\boldmath $z, dx, dy$}$ & \bf{171.4}   & \bf{1.35} & \bf{299.5} & $\mbox{\boldmath $7.513 \pm 0.082$}$ & $\mbox{\boldmath $14.86 ^{+0.11}_{-0.44}$}$ & $\mbox{\boldmath $0.00 \pm 0.24$}$& $\mbox{\boldmath $0.29971 \pm 0.00022$}$\\[3pt]
$z, t, dx, dy$ & 176.3 & 1.35 & 299.5 & 7.483 $\pm$ 0.082 & 14.87 $^{+0.12}_{-0.44}$ & 0.00 $\pm$ 0.24 &  0.29974 $\pm$ 0.00022\\[3pt]
$t, t^2, dx, dy$ & 176.4 & 1.35 & 299.9 & 7.432 $\pm$ 0.083 & 14.88 $^{+0.11}_{-0.44}$  & 0.00 $\pm$ 0.24 & 0.29973 $\pm$ 0.00022
\enddata
\tablenotetext{a}{\ The fitting results for the adopted baseline model are indicated as bold text.}
\tablenotetext{b}{\ The subscript ``all'' denotes that the entire light curve is used for the $\chi^2$ calculation.}
\tablenotetext{c}{\ The uncertainties are calculated after the flux errors have been rescaled so that the reduced $\chi^2_\mathrm{all}$ value is unity.} 
\tablenotetext{d}{\ The $b$ value is allowed to be minus in the fitting process, but  the absolute value is used for modeling a transit light curve.}
\tablenotetext{}{}
\end{deluxetable*}

In order to select an appropriate aperture radius,  we create a number of trial light curves by changing the aperture radius by one pixel, and evaluate the dispersion of the out-of-transit (OOT) part of these trial light curves.
We find that the light curve produced with an aperture radius of 34 pixels gives the minimum dispersion, and therefore use this light curve for further analyses. The selected light curve is shown in Figure \ref{fig:rawlc_isle}.
Subsequently, we correct the systematic trend in the baseline of the light cure, which is apparent in the OOT part of the light curve. Generally, such a trend could arise from, e.g., air mass change, slow variability of the target and/or comparison stars themselves, and stellar displacement on the detector.
One approach to correct this trend is to simultaneously fit the in-transit and OOT parts of the light curve with a transit + baseline-correction function.
This approach has an advantage of estimating reasonable uncertainties of the transit-model parameters by taking into account the correlations between the transit-model parameters and those of the baseline function. 
However, generally speaking, which formula is appropriate for a baseline function is not obvious, and if an inappropriate baseline function is chosen, then there is the possibility that the in-transit data affect on the baseline-function parameters and thereby the transit is deepened or shallowed systematically. 
In order to avoid this effect, we adopt the following approach: we first determine a baseline-correction function and its coefficients by using the OOT light curve alone, and then adopt this function to correct the entire light curve. 

\begin{figure}
\begin{center}
\includegraphics[width=8.5cm]{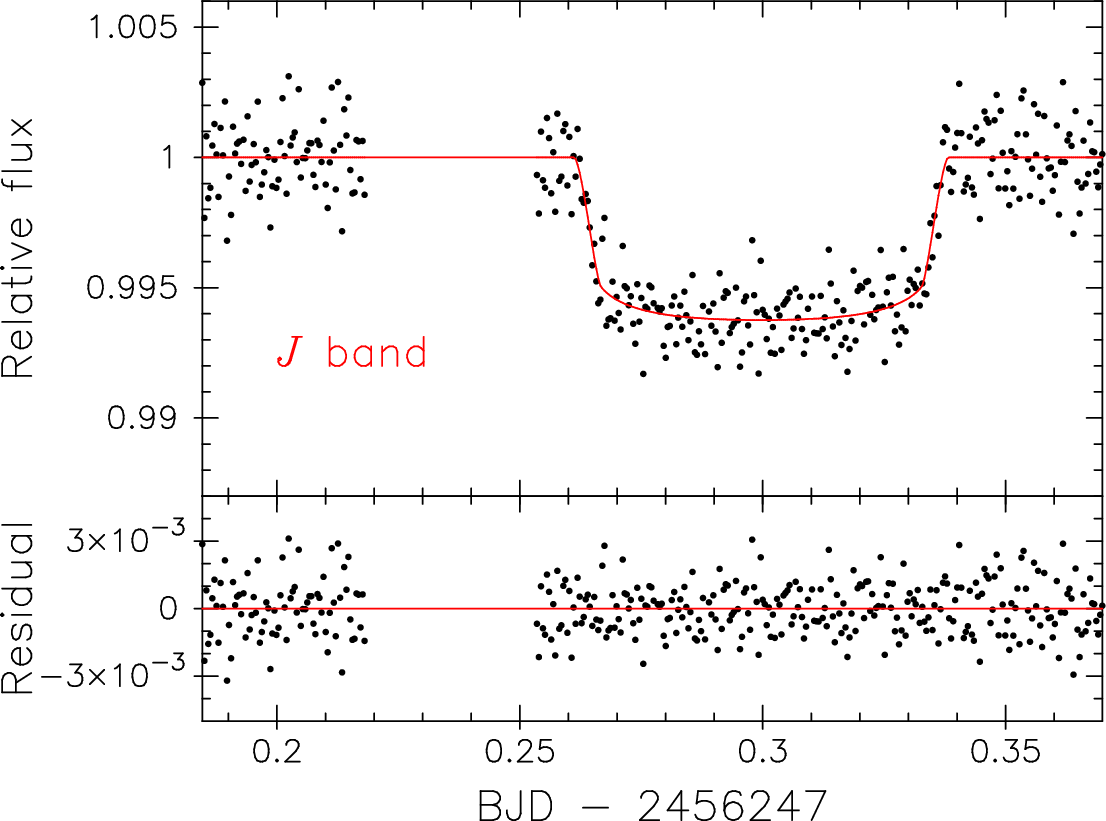}
\caption{Top panel: the baseline-corrected $J$-band light curve. The best-fit transit model derived in Section \ref{sec:reduction_isle} is shown as solid line. Bottom panel: the residual light curve. The rms of the residuals is  $1.18 \times 10^{-3}$. \label{fig:fittedlc_isle}}
\end{center}
\end{figure}

More specifically, we use the following procedure. First, we assume that the baseline function can be expressed in the following formulae:
\begin{eqnarray}
F_\mathrm{base} &=& k_0 \times 10^{-0.4 \Delta m_\mathrm{corr}},\\
\Delta m_\mathrm{corr} &=& \sum_{i=1} k_i X_i,
\end{eqnarray}
where $F_\mathrm{base}$ is the baseline flux, $\{{\bf X}\}$ are variables, and $\{{\bf k}\}$ are coefficients.
For the variables $\{{\bf X}\}$, we test several combinations of $z$, $t$, $t^2$, $dx$, and $dy$, where  
$z$ is air mass, $t$ is time, $dx$ and $dy$ are the relative centroid positions in the $x$- and $y$-directions, respectively.
Next, we evaluate the Bayesian information criteria \citep[BIC;][]{1978Schwarz} for the respective baseline models.
The BIC value for the OOT light curve  is given by $\mathrm{BIC}_\mathrm{oot} \equiv \chi_\mathrm{oot}^2 + k \ln N_\mathrm{oot}$, where $\chi_\mathrm{oot}^2$ is the $\chi^2$ value of the baseline fit for the OOT part, $k$ is the number of free parameters, and $N_\mathrm{oot}=155$ is the number of OOT data points. 
We find that the $\{{\bf X}\} = \{z, dx, dy\}$ model gives the minimum BIC$_\mathrm{oot}$ value among all the baseline models, and therefore apply this model to correct the systematic trend in the entire light curve. 
We note that this result is robust over other trial light curves produced with slightly different aperture radii rather than 34 pixels.
In Table \ref{tbl:amoeba}, we list the BIC$_\mathrm{oot}$ values and the rms of the OOT light curve (rms$_\mathrm{oot}$) for the four representative baseline models of  $\{X\}$ = $\{z\}$, $\{z, dx, dy\}$, $\{z, t, dx, dy\}$, and $\{t, t^2, dx, dy\}$.
Noticeably,  including $dx$ and $dy$ in $\{{\bf X}\}$ provides a significant improvement on both the BIC$_\mathrm{oot}$ and rms$_\mathrm{oot}$ values, from BIC$_\mathrm{oot}=224.2$ and rms$_\mathrm{oot}$ $=1.61\times10^{-3}$  for the $\{z\}$ model to BIC$_\mathrm{oot}=171.4$ and rms$_\mathrm{oot}$ $=1.35\times10^{-3}$ for the $\{z, dx, dy\}$ model. This fact indicates that the stellar displacements on the detector produce significant systematics on the photometry, and the displacements during our observations were sufficiently suppressed such that the systematics can be corrected, owing to the development of the offset-correction system (Section \ref{sec:isle_obs}).
In Figure \ref{fig:rawlc_isle}, we also plot the best-fit OOT light curve with a baseline model of $\{{\bf X}\} = \{z, dx, dy\}$, along with the time variations of $z$, $dx$, and $dy$. The corrected light curve is shown in Figure \ref{fig:fittedlc_isle}, and its numerical data are reported in Table \ref{tbl:sample_lc}.

\begin{deluxetable}{lcc}
\tablewidth{7.5cm}
\tablecaption{$J$, $I_\mathrm{c}$, $R_\mathrm{c}$, and $g'$ Light Curves \label{tbl:sample_lc}}
\tablehead{
\colhead{BJD$_\mathrm{TDB}$ (-2450000)} & \colhead{Flux} & \colhead{$\sigma$}\\
\colhead{(days)} & &
}
\startdata
\multicolumn{3}{c}{$J$ band}\\[3pt]
\hline\\[-5pt]
6247.184623 &1.002867 &0.001560\\[3pt]
6247.185040&0.997683&0.001549\\[3pt]
6247.185456&1.000811&0.001548\\[3pt]
6247.185873&0.998436&0.001542\\[3pt]
6247.186289&0.998839&0.001535
\enddata
\tablecomments{All fluxes are corrected for baseline trends.\\
(This table is available in its entirety in a machine-readable form in the online version of the {\it Astrophysical Journal}. A portion is shown here for guidance regarding its form and content.)}
\end{deluxetable}

The selected baseline model of $\{{\bf X}\} = \{z, dx, dy\}$, however, is not so significantly favored compared to other baseline models of $\{t, z, dx, dy\}$ and  $\{t, t^2, dx, dy\}$, given a BIC$_\mathrm{oot}$ difference of only $\sim5$.
Therefore, the choice of the baseline model could be a cause of a systematic error on the final result, i.e., transit parameters.
In order to evaluate the impact of this systematic effect, we fit the entire light curves corrected by different baseline models with a transit model and compare the resultant parameters.
The transit parameters we use are the planet-star radius ratio, $R_p/R_s$, the mid transit time, $T_\mathrm{c}$, the semi-major axis normalized by the stellar radius, $a/R_s$, the impact parameter, $b \equiv a \cos i/R_s$, where $i$ is the orbital inclination, and the coefficients for a stellar limb-darkening effect. We adopt the orbital eccentricity and  orbital period as $e=0$ and $P = 3.33665$ days, respectively, from D13.
  For a stellar limb-darkening model, we use the quadratic limb-darkening raw, $I(\mu) =  1 - u_1 (1-\mu) - u_2 (1-\mu)^2$, where $I$ is the intensity, $\mu$ is the cosine of the angle between the line of sight and the line from the stellar center to the position of the stellar surface, and $u_1$ and $u_2$ are the coefficients. Because $u_1$ and $u_2$ are heavily correlated and cannot independently be constrained well, we let $u_1$ free while fix $u_2$ at a theoretical value during the fitting process; we adopt $u_2 = 0.255$, witch is the mean of the two values for  \{$T_\mathrm{eff}$, $\log g_s$\} = \{3600, 4.5\} and \{3600, 5.0\}, where $T_\mathrm{eff}$ (K) is the stellar effective temperature and $\log g_s$ (cgs) is the stellar surface gravity, in the table given by \citet{2012A&A...546A..14C}. (The $T_\mathrm{eff}$ and $\log g_s$ values for GJ3470 were derived by D13 as $3600\pm100$ and $4.658\pm0.035$, respectively.)
For calculating a transit model, we use the analytic formula given by \citet{2009ApJ...690....1O}, which is equivalent to that of \citet{2002ApJ...580L.171M} when using the quadratic limb-darkening raw. The best-fit parameters are determined by the AMOEBA algorithm \citep{1992nrca.book.....P}. We iteratively fit the data by eliminating $>$3-$\sigma$ outliers, resulting in one data point being removed. Before estimating 1-$\sigma$ uncertainties of the fitted parameters, the photometric errors are rescaled so that the reduced $\chi^2$ value for the best-fit transit model becomes unity.
 
In Figure \ref{fig:fittedlc_isle}, the best-fit transit model for the light curve corrected with the $\{{\bf X}\}$ = $\{z, dx, dy\}$ model is plotted. 
The derived best-fit values and uncertainties of $R_p/R_s$, $a/R_s$, and $T_\mathrm{c}$, as well as the $\chi^2$ values are listed in Table \ref{tbl:amoeba}.
While the $\chi^2$ value for the  $\{{\bf X}\}$ = $\{z\}$ model is relatively large as expected, those for the other three models of $\{z, dx, dy\}$, $\{z, t, dx, dy\}$, and $\{t, t^2, dx, dy\}$ are very close to each other.
However, the $R_p/R_s$ values for these three models are slightly different, with the largest difference of 0.00081 between $\{z, dx, dy\}$ and $\{t, t^2, dx, dy\}$, which is comparable to the 1-$\sigma$ uncertainties of $R_p/R_s$.
Therefore, we consider this discrepancy to be a systematic error in $R_p/R_s$ and take it into account in the study of the wavelength dependency of $R_p/R_s$ discussed in Section \ref{sec:radius_ratio}.
 On the other hand, the baseline-model dependences on $a/R_s$, $b$, and $T_\mathrm{c}$ are negligible compared to their 1-$\sigma$ uncertainties.

In addition, in order to estimate the significance of time-correlated (red) noise in the light curve, which can arise from factors such as short-term stellar variability  and changing atmospheric conditions \citep{2006MNRAS.373..231P}, we estimate the red noise factor $\beta$ by using the ''time-average" method according to \citet{2008ApJ...683.1076W}. Namely, first, the residual light curve is binned into $M$ bins by averaging every $N$ data point, in order to calculate the standard deviation of the $M$ bins, $\sigma_{N, \mathrm{obs}}$. Subsequently, $\beta$ is calculated by dividing $\sigma_{N, \mathrm{obs}}$ by $\sigma_{N, \mathrm{exp}} \equiv \sigma_1  \sqrt{M/N(M-1)}$, where $\sigma_1$ is the standard deviation of the  unbinned residual light curve. As a result, we find that $\beta$ does not exceed unity for $N=4$ to 15, indicating that the red noise is negligible for the $J$-band light curve.

\subsection{Reduction and Baseline Correction for the MITSuME data}
\label{sec:reduction_mitsume}

The obtained MITSuME images are reduced in the same way as the ISLE images.
For the flat-fielding correction, 27 twilight-flat images for each band obtained on the observing night are used to create each flat-fielding image.
We then apply aperture photometry for GJ3470 and $\sim$25 of bright stars, all of which are brighter than 15 mag in the respective bands, 
with a number of trial aperture radii incremented by 0.5 pixels.
For a trial aperture radius, we create a number of trial light curves by dividing the target flux by each sum of the fluxes of a trial combination of comparison stars among the $\sim$25 comparison stars.
Among a number of combinations of aperture radii and comparison stars,
we first visually select several good combinations that produce less-dispersed light curves with respect to a transit signature.
Next, we fit each trial light curve including the transit part with a tentative transit+baseline model as described below, in order to select the least-dispersed light curve for each band. We note that the reason we include the transit part  this time is that, unlike the $J$-band data, MITSuME light curves do not have enough OOT data points due to the longer exposure time and the relatively short coverage of the OOT part before the transit.

\begin{deluxetable}{lccc}
\tabletypesize{\footnotesize}
\tablecaption{Fitting results for different baseline models\label{tbl:amoeba_oao50} \tablenotemark{a}}
\tablewidth{8cm}
\tablehead{
\colhead{Variables} & \colhead{BIC} & \colhead{rms} & \colhead{$R_p/R_s$}\\
 \colhead{$\{{\bf X}\}$} & & \colhead{($\times 10^{-3}$)}  & \colhead{($\times 10^{-2}$)}
}
\startdata
\multicolumn{4}{c}{MITSuME $I_\mathrm{c}$}\\[3pt]
$\bf{z}$ & \bf{185.5} & \bf{1.81} & $\mbox{\boldmath $7.97 \pm 0.13$}$ \\[3pt]
$z, t$ & 192.1 & 1.82 & $8.06 \pm 0.12$\\[3pt]
$t, t^2$ & 191.8 & 1.82 & $8.11 \pm 0.12$\\[3pt]
\hline\\[-5pt]
\multicolumn{4}{c}{MITSuME $R_\mathrm{c}$}\\[3pt]
$z$ & 186.5 & 2.50 & $8.32 \pm 0.17$   \\[3pt]
$\bf{z, t}$ & \bf{186.5} & \bf{2.46} & \mbox{\boldmath $7.72 \pm 0.17$} \\[3pt]
$t, t^2$ &186.5 & 2.46 & $7.74 \pm 0.17$ \\[3pt]
\hline\\[-5pt]
\multicolumn{4}{c}{MITSuME $g'$}\\[3pt]
$\bf{z}$ &\bf{184.5} & \bf{5.12} & \mbox{\boldmath $8.16 \pm 0.30$} \\[3pt]
$z, t$ & 189.6 & 5.12 & $8.08 \pm 0.30$ \\[3pt]
$t, t^2$ & 189.7 & 5.12 & $8.07 \pm 0.30$
\enddata
\tablenotetext{a}{\ The fitting results for the adopted baseline models are indicated as bold text.}
\tablenotetext{}{}
\end{deluxetable}

For the tentative baseline model, we use Equation (1) and (2) with $\{{\bf X}\}$ = $\{z\}$. 
As for the tentative transit model, we use the same parameterization as in Section \ref{sec:reduction_isle}, and force $a/R_s$ and $b$ to the values derived from D13, namely 13.4 and 0.40, respectively. This is because the $Spitzer$ light curves used in D13 are more precise than those of MITSuME, and we are more focusing on studying  the wavelength dependence of $R_p/R_s$ (see Section \ref{sec:radius_ratio}) rather than constraining other parameters. 
For the $I_\mathrm{c}$- and $R_\mathrm{c}$-band light curves,  we fix $u_2$ at the theoretical values of 0.338 and 0.322, respectively, while we leave $u_1$ free during the fitting process. For the $g'$-band light curve, we fix both $u_1$ and $u_2$ at the theoretical values of 0.486 and 0.289, respectively, because this light curve is too poor to constrain $u_1$ or $u_2$. These theoretical values are derived in the same way as for the $J$-band light curve.

As a result,  we select the least-dispersed light curves that are produced with the aperture radii of 7.0, 8.0, and 6.0 pixels, and the number of comparison stars of 4, 10, and 8, for $I_\mathrm{c}$, $R_\mathrm{c}$, and $g'$ bands, respectively. 
We summarize the photometric properties of the selected comparison stars in Table \ref{tbl:list_compstars}.
We note that these results are robust over different baseline models.

Subsequently, in order to select appropriate baseline models for correcting systematic trends,
we again fit the selected light curves with transit + several baseline models.
At this time, we  test the three baseline models of $\{{\bf X}\} = \{z\}$, $\{z, t\}$, and $\{t, t^2\}$.
We do not include $dx$ and $dy$ in  $\{{\bf X}\}$ because we do not find any improvement for any bands when we preliminarily fit the light curves by including $dx$ and $dy$.
In Table \ref{tbl:amoeba_oao50}, we list the resultant BIC and rms values for the three baseline models for each band.
For the $I_\mathrm{c}$ and $g'$ bands, we find that the baseline model of $\{{\bf X}\} = \{z\}$ gives the minimum BIC values, and therefore adopt it for correcting the systematic trends in these light curves.
On the other hand, for the $R_\mathrm{c}$-band data, all three baseline models give almost the same BIC values, indicating that the three models have the same statistical significance. So, we decide to select the model $\{{\bf X}\} = \{z, t\}$ among the three models, based on the fact that the rms values for $\{z, t\}$ and $\{t, t^2\}$ ($2.46 \times 10^{-3}$) are slightly better than that for $\{z\}$  ($2.50 \times 10^{-3}$), and it is physically more straightforward to include air mass in the variables rather than just a polynomial function of time.

As also discussed in Section \ref{sec:reduction_isle}, the selection of baseline models could affect the final result. 
In Table \ref{tbl:amoeba_oao50}, we also list the resultant $R_p/R_s$ values for the respective models for each band.
The largest differences of $R_p/R_s$ are  0.0014, 0.0060, and 0.009, for the $I_\mathrm{c}$, $R_\mathrm{c}$, and $g'$ bands, respectively.
The differences for the $I_\mathrm{c}$ and $g'$ bands  are comparable to and less than the 1-$\sigma$ uncertainties of $R_p/R_s$, respectively, while that for the $R_\mathrm{c}$ band is more than 3 times the 1-$\sigma$ uncertainty of $R_p/R_s$. 
These possible systematic offsets will be considered in the discussion on the wavelength dependence of $R_p/R_s$ in Section \ref{sec:radius_ratio}.

\begin{figure}
\begin{center}
\includegraphics[width=8.5cm]{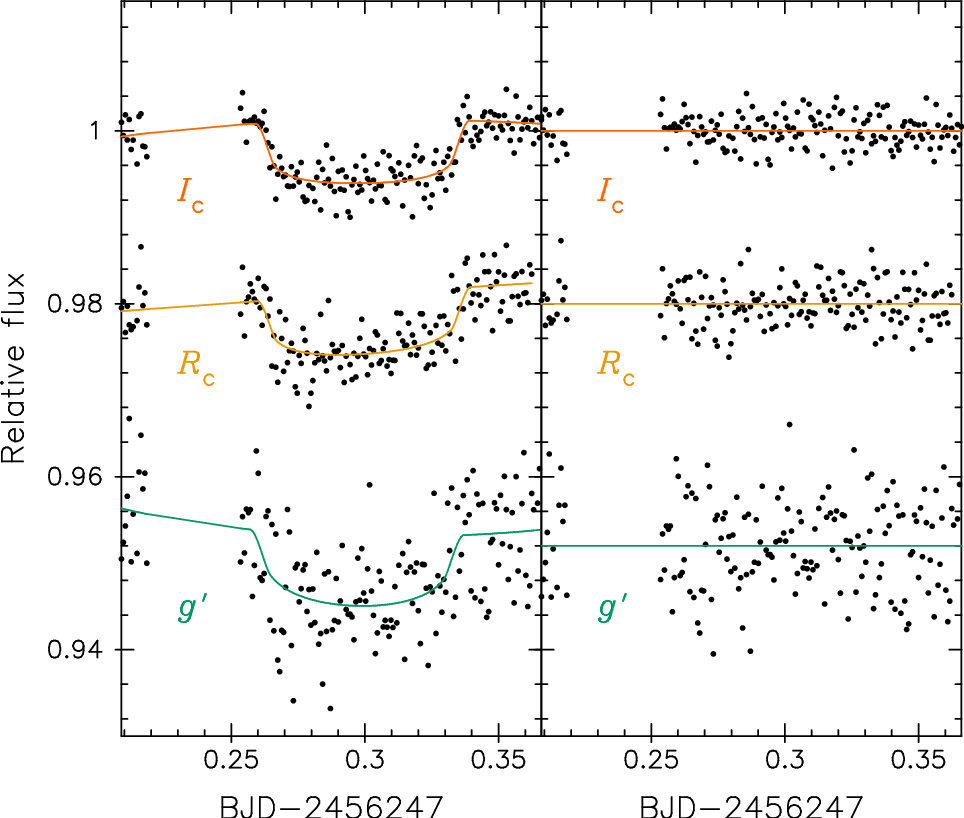}
\caption{Left panel: the uncorrected light curves of MITSuME $I_\mathrm{c}$, $R_\mathrm{c}$, and $g'$ bands, from top to bottom. The best-fit baseline models are over plotted as solid lines. For display, -0.02 and -0.048 are added in the $R_\mathrm{c}$ and $g'$ light curves, respectively. Right panel: the same as the left panel but residual light curves. The rms of these residuals are $1.78 \times 10^{-3}$, $2.45 \times 10^{-3}$, and $5.27 \times 10^{-3}$, for $I_\mathrm{c}$, $R_\mathrm{c}$, and $g'$ bands, respectively. \label{fig:fittedlc_oao50}}
\end{center}
\end{figure}

In Figure \ref{fig:fittedlc_oao50}, we show the uncorrected light curves and the best-fit transit+baseline models of the three bands, as well as their residuals. After the baseline for each light curve is corrected, error bars are rescaled so that the reduced $\chi^2$ for a transit-model fit becomes unity. 
The numerical data of the corrected light curves are compiled in Table \ref{tbl:sample_lc}.
In addition, we estimate the red noise factor $\beta$ for each light curve in the same way as in Section \ref{sec:reduction_isle}, and find that $\beta$ is unity for almost all cases of $N=4$ to 15 for all bands.

\section{ANALYSIS}
\label{sec:mcmc}
In order to derive the final transit parameters and their uncertainties, we analyze the corrected $J$-, $I_\mathrm{c}$-, $R_\mathrm{c}$-, and $g'$-band light curves simultaneously with the Markov Chain Monte Carlo (MCMC) method by using a customized code \citep{2012arXiv1210.3169N}. 
In this analysis, we treat $b$, $a/R_s$, and $T_\mathrm{c}$ as common parameters for all four light curves, while $R_p/R_s$, $u_1$, and $u_2$ are treated as independent parameters for the respective light curves so as to take the wavelength dependences of these parameters into account. The values of $b$, $a/R_s$, $T_\mathrm{c}$, and $R_p/R_s$ for each band are left free during the MCMC process.  The $u_1$ and $u_2$ values are treated the same way as previously (see Section \ref{sec:reduction_isle} and \ref{sec:reduction_mitsume}); $u_1$ for the $J$, $I_\mathrm{c}$, and $R_\mathrm{c}$ bands is left free, while $u_1$ for the $g'$ band and $u_2$ for all the bands are fixed at the theoretical values. We  adopt $e=0$ and $P=3.33665$ days in the same way as before.

\begin{deluxetable*}{lccc}
\tabletypesize{\footnotesize}
\tablewidth{17cm}
\tablecaption{MCMC results for transit parameters\label{tbl:mcmc}}
\tablehead{
\colhead{Parameter} & \colhead{D13} &\colhead{This Work (without $b$ prior)} & \colhead{This Work (with $b$ prior) } 
}
\startdata
Impact parameter, $b$ $(\equiv a \cos i / R_s)$ & $0.40^{+0.06}_{-0.08}$ & $<$ 0.281 (1-$\sigma$ upper limit) & $0.337^{+0.067}_{-0.070}$\\[3pt]
Scaled semi-major axis, $a/R_s$ & $13.42^{+0.55}_{-0.53}$ & 14.70 $^{+0.17}_{-0.43}$ & 14.02 $^{+0.33}_{-0.39}$\\[3pt]
Mid-transit time, $T_c$ [BJD$_\mathrm{TDB}$ - 2450000]& $6090.47705 \pm 0.00014$ & 6247.29954 $\pm$ 0.00019 & 6247.29951 $\pm$ 0.00020\\
& 6093.81372 $\pm$ 0.00015 & & \\[5pt]
Planet-to-star radius ratio & \\[1pt]
$R_p/R_{s}$ ($4.5\mu$m) & $0.07798^{+0.00046}_{-0.00045}$ & ... & ... \\[3pt]
$R_p/R_{s}$ ($J$) & ... & 0.07536 $\pm$ 0.00079 & 0.07577 $^{+0.00072}_{-0.00075}$\\[3pt]
$R_p/R_{s}$ ($I_c$) & ... & 0.0797 $\pm$ 0.0014 & 0.0802 $\pm$ 0.0013\\[3pt]
$R_p/R_{s}$ ($R_c$) & ... & 0.0770 $\pm$ 0.0019 & 0.0776 $\pm$ 0.0018\\[3pt]
$R_p/R_{s}$ ($g'$) & ... & 0.0800 $\pm$ 0.0030 & 0.0809 $\pm$ 0.0031\\[5pt]
Limb-darkening coefficients & \\[1pt]
$u_1$ ($J$) & ... & 0.149 $\pm$ 0.074 & 0.137 $^{+0.077}_{-0.073}$\\[3pt]
$u_2$ ($J$) & ... & 0.255 (fixed) & 0.255 (fixed)\\[3pt]
$u_1$ ($I_c$) & ... & 0.20 $\pm$ 0.11 & 0.19 $\pm$ 0.11\\[3pt]
$u_2$ ($I_c$) & ... & 0.338 (fixed) & 0.338 (fixed)\\[3pt]
$u_1$ ($R_c$) & ... & 0.26 $\pm$ 0.15 & 0.25 $\pm$ 0.15\\[3pt]
$u_2$ ($R_c$) & ... & 0.322 (fixed) & 0.322 (fixed)\\[3pt]
$u_1$ ($g'$) &  ... & 0.486 (fixed) &  0.486 (fixed)\\[3pt]
$u_2$ ($g'$) & ... & 0.289 (fixed) & 0.289 (fixed)
\enddata
\end{deluxetable*}

\begin{figure*}
\begin{center}
\includegraphics[width=12cm]{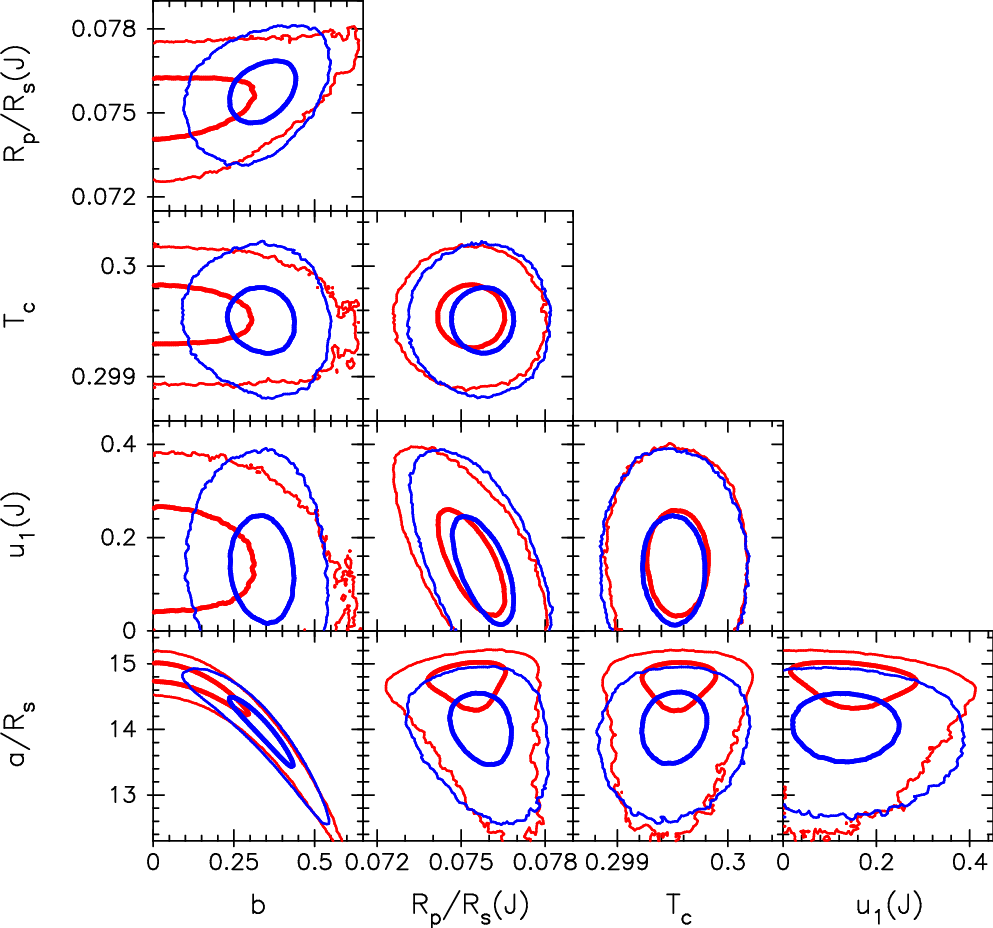}
\caption{Two-dimensional 68.3\% (thick lines) and 99.7\% (thin lines) confidence regions for selected parameters calculated from the posterior probability distributions as a consequence of the MCMC analysis. The red contours show the results from the MCMC analysis without any prior, while the blue contours indicate those with a $b$ prior.
\label{fig:correlationmap1}}
\end{center}
\end{figure*}

\begin{figure}
\begin{center}
\includegraphics[width=8.5cm]{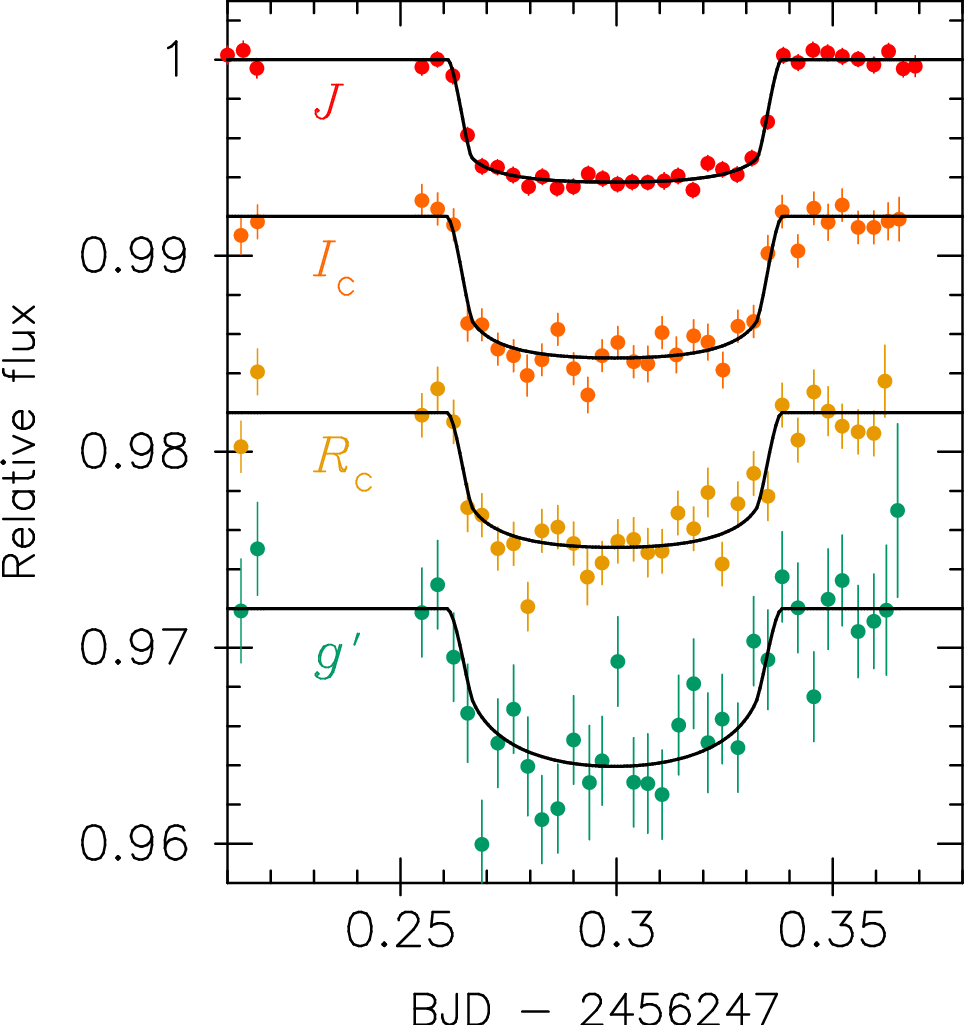}
\caption{Overall light curves after the baseline correction, the error normalization, and five minute binning. The $J$-, $I_\mathrm{c}$-, $R_\mathrm{c}$-, and $g'$-band light curves are shown from top to bottom. The offsets of -0.08, -0.18, and -0.28 are added for the $I_\mathrm{c}$-, $R_\mathrm{c}$-, and $g'$-band light curves, respectively, for clarify. The best-estimated transit models derived from the MCMC analysis without $b$ prior are shown as solid lines. \label{fig:lc_all}}
\end{center}
\end{figure}

In an MCMC chain, we start with a set of parameters that provides the minimum $\chi^2$ value determined by the AMOEBA algorithm. 
The $\chi^2$ value is given by
\begin{eqnarray}
\label{eq:chisq1}
\chi^2 = \sum_i \left( \frac{f_{\mathrm{obs}, i} - f_\mathrm{model}}{\sigma_i} \right)^2,
\end{eqnarray}
where $f_{\mathrm{obs}, i}$ is the $i$-th photometric flux, $\sigma_i$ is the $i$-th photometric error, and $f_\mathrm{model}$ is the model flux calculated from the given parameter set.    
Then, the next parameter set is randomly selected, where we assume Gaussian distributions whose means are the respective current values, and standard deviations are proportional to their 1-$\sigma$ errors. This parameter set is always accepted when it gives $\Delta \chi^2 < 0$, where $\Delta \chi^2$ is the $\chi^2$ difference between the next parameter set and the current one; otherwise, it is accepted according to an acceptance probability which is given by $p = \exp^{\Delta \chi^2 / 2}$.
This process (link) is repeated for 10$^6$ times in a chain, from which the first 10$^5$ links are removed as a burn-in portion.
The widths of the Gaussian distributions for jumping parameters are set so that the acceptance ratio becomes about 25\%.
We run 10 independent chains and create merged posterior distributions of the respective parameters. 
We define 1-$\sigma$ statistical uncertainties as the range of parameters between 15.87\% and 84.13\% of the posterior distributions.  
The resultant values and uncertainties are listed in Table \ref{tbl:mcmc}, and two-dimensional 68.3\% and 99.7\% confidence regions for  selected parameters drawn from the posterior distributions are shown as red contours in Figure \ref{fig:correlationmap1}. The derived model light curves for the respective bands are shown in Figure \ref{fig:lc_all}, along with the observed data binned in 5-minute intervals.

We confirm that the values of $b$, $a/R_s$, and $R_p/R_s$ for the $I_\mathrm{c}$, $R_\mathrm{c}$, and $g'$ bands are all consistent with those reported by D13 within 2 $\sigma$.
On the other hand, we find that the $R_p/R_s$ value for the $J$-band data, $0.07536 \pm 0.00079$, is inconsistent with that for the $I_\mathrm{c}$-band data of $0.0797 \pm 0.0014$ by 2.6 $\sigma$, and is also inconsistent with that for the $Spitzer$'s 4.5-$\mu$m data of $0.07798^{+0.00046}_{-0.00045}$ by 2.9 $\sigma$. 
One possible cause of these discrepancies is a systematic effect originated from stellar activity \citep[e.g.][]{2009A&A...505.1277C}. Occulting cool starspots by a transiting planet can induce a bump in the transit light curve, easily affecting the fitted parameters. Although no such feature is identified in the light curves obtained by this work or by $Spitzer$, there is still the possibility that the planet occulted cool spots existing near the stellar limb, which would be difficult to identify. In this case, the transit parameters such as $b$ and $a/R_s$ would easily be affected. The $R_p/R_s$ value would  also be affected, because $R_p/R_s$ and $b$ (or $a/R_s$) are weakly correlated (see Figure \ref{fig:correlationmap1}). This might be the case, because the $b$ value derived from our data ($< 0.281$ as 1-$\sigma$ upper limit) is inconsistent with that by D13 ($0.40^{+0.06}_{-0.08}$) by more than 1 $\sigma$. In order to reduce such a systematical effect on $R_p/R_s$, we repeat the MCMC analysis by using the $b$ value reported by D13 as prior information.
In this analysis,   we use the following function instead of Equation (\ref{eq:chisq1}) for calculating the $\chi^2$ value:
\begin{eqnarray}
\label{eq:chisq2}
\chi^2 = \sum_i \left\{ \left( \frac{f_{\mathrm{obs}, i} - f_\mathrm{model}}{\sigma_i} \right)^2  \right\} + \left( \frac{ b - b_\mathrm{prior} }{ \sigma_{b_\mathrm{prior}} } \right)^2,
\end{eqnarray}
where we adopt $b_\mathrm{prior}$ = 0.40, and $\sigma_{b_\mathrm{prior}}$ = 0.06 if ($b - b_\mathrm{prior}$) $> 0$; otherwise $\sigma_{b_\mathrm{prior}}$ = 0.08. 
This formula imposes a penalty on the $\chi^2$ value if $b$ deviates from the prior value during the MCMC process.
The resultant values and uncertainties are summarized in Table \ref{tbl:mcmc}, and the two-dimensional correlation maps for selected parameters are plotted as blue contours in Figure \ref{fig:correlationmap1}. As a result, we find that our $b$ and $a/R_s$ values become consistent with those by D13 within 1 $\sigma$. However,  although the $R_p/R_s$ values  for our data ($J$, $I_\mathrm{c}$, $R_\mathrm{c}$, and $g'$ bands) become closer to that for the 4.5-$\mu$m band, there is still disagreement between $J$ and $I_\mathrm{c}$ by 2.5 $\sigma$, and between $J$ and  4.5 $\mu$m by 2.6 $\sigma$.
As discussed in Section \ref{sec:radius_ratio}, these discrepancy can likely be attributed to the wavelength dependence of the planetary atmospheric opacity.

\section{DISCUSSION}
\label{sec:discussion}

\subsection{Physical Parameters of the Planetary System}
\label{sec:phys_param}
In this section, we focus on the physical parameters of the planetary system.
One of our goals in this section is to test the very low density of GJ3470b suggested by D13 with our independent observations, and therefore we here discuss the results of the MCMC analysis without $b$ prior.

\begin{deluxetable*}{lccc}
\tabletypesize{\footnotesize}
\tablewidth{17cm}
\tablecaption{Physical parameters\label{tbl:mc}}
\tablehead{
\colhead{Parameter} & \colhead{D13} & \colhead{This work (without $b$ prior)} & \colhead{This work (with $b$ prior)}
}
\startdata
{\it Stellar parameters}\\[2pt]
Stellar mass, M$_s$ (M$_\odot$) & $0.539^{+0.047}_{-0.043}$ &  $0.557^{+0.028}_{-0.020}$ &  $0.594^{+0.029}_{-0.026}$\\[3pt]
Stellar radius, R$_s$ (R$_\odot$) & $0.568^{+0.037}_{-0.031}$ &  $0.526^{+0.023}_{-0.011}$ & $0.563^{+0.024}_{-0.020}$\\[3pt]
Stellar density, $\rho_s$ ($\rho_\odot$) & $2.91^{+0.37}_{-0.33}$ & $3.83^{+0.14}_{-0.32}$ & $3.32^{+0.24}_{-0.27}$\\[3pt]
Stellar surface gravity, $\log g_s$ (cgs) & $4.658 \pm 0.035$ &  $4.741^{+0.009}_{-0.019}$ & $4.710^{+0.016}_{-0.019}$\\[3pt]
\hline\\[-5pt]
{\it Planetary parameters}\\[2pt]
Orbital period, $P$ (days) & $3.33665 \pm 0.00005$ &  $3.336649 \pm 0.000005$ & $3.336648 \pm 0.000005$\\[3pt]
Semi-major axis, $a$ (AU)  & $0.03557^{+0.00096}_{-0.00100}$ & $0.03596^{+0.00059}_{-0.00044}$ & $0.03674^{+0.00059}_{-0.00054}$\\[3pt]
Planetary mass, M$_p$ (M$_\oplus$) & $13.9^{+1.5}_{-1.4}$ &  14.1 $\pm$ 1.3 & 14.6 $\pm$ 1.4\\[3pt]
Planetary radius, R$_p$ (R$_\oplus$) & $4.83^{+0.22}_{-0.21}$ & $4.32^{+0.21}_{-0.10}$ & $4.65^{+0.22}_{-0.18}$\\[3pt]
Planetary density, $\rho_p$ (g cm$^{-1}$) & $0.72^{+0.13}_{-0.12}$ & 0.94 $\pm$ 0.12 & 0.80 $\pm$ 0.11\\[3pt]
Planetary surface gravity, $g_p$ (m s$^{-2}$) & $5.75^{+0.85}_{-0.86}$ & $7.25^{+0.75}_{-0.78}$ & $6.58^{+0.72}_{-0.70}$
\enddata
\end{deluxetable*}

First, we refine the orbital period by a linearly fit to our $T_\mathrm{c}$ value and the ones reported by D13 for two continuous transits. The resultant orbital period is $P = 3.336649 \pm 0.000005$ days. This is in good agreement with $P=3.33665 \pm 0.00005$ days, which was derived by D13 based on the two transits observed by $Spitzer$ and four transits from \citet{2012A&A...546A..27B}, with a difference of only $0.09 \pm 4.34$ s. 
In addition, the difference between our measured $T_\mathrm{c}$ value and the expected one calculated from the ephemeris provided by D13 is $2 \pm 204$ sec, indicating that no TTV is observed.

\begin{figure}
\begin{center}
\includegraphics[width=8.5cm]{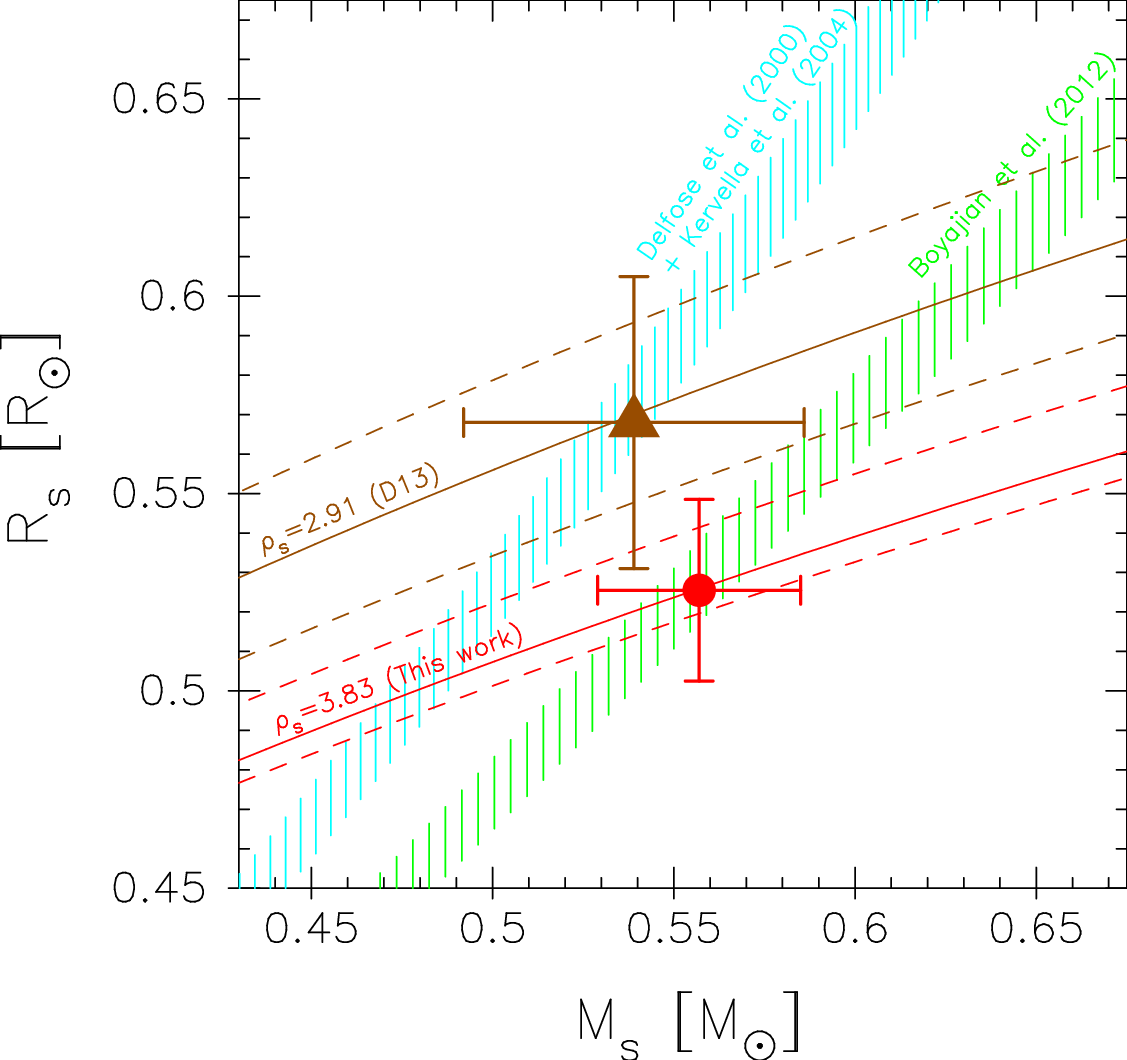}
\caption{ Comparison of $M_s$ and $R_s$ calibrations between this work and D13. The stellar density derived in this work (MCMC analysis without $b$ prior) and that by D13 are indicated by lower (red in online) and upper (brown in online) lines. Solid and dashed lines represent median and 1 $\sigma$ uncertainties, respectively. The mass-radius relation we adopted (Equation (4) in  \citet{2012ApJ...757..112B}) is shown as the right (green in online) shaded region (1-$\sigma$ region), while the mass-radius relation for GJ3470 drawn by connecting a mass-luminosity \citep{2000A&A...364..217D} and a diameter-color-luminosity \citep{2004A&A...426..297K} relation is shown as the left (light-blue in online) shaded region, where 2\% uncertainty in $R_s$ is adopted. 
The $M_s$ and $R_s$ values derived in this work and  by D13 are indicated as circle and triangle, respectively. \label{fig:MsRs}}
\end{center}
\end{figure}

Next, we investigate the stellar mass and radius.
The mean stellar density, $\rho_s$, can be directly derived from $a/R_s$ and $P$ via the following relation assuming a circular orbit: $\rho_s = 0.01342 \times (a/R_s)^3 / (P (\mathrm{days}))^2$ $\rho_\odot$. From our MCMC results, we derive the stellar density to be $\rho_s = 3.83^{+0.14}_{-0.32}$ $\rho_\odot$, which is broadly consistent with the value derived by D13 ($2.91^{+0.37}_{-0.33}$ $\rho_\odot$) within 2 $\sigma$, although
our value is 32\% larger than that by D13. 
The stellar mass and radius can be derived from the stellar density, combined with one more piece of information such as a stellar color (or temperature), luminosity (i.e., apparent brightness and distance to the star), and/or a stellar mass-radius relation.
Because a trigonometric parallax for GJ3470 has not been measured so far, D13 solved for the stellar mass, radius, and parallax simultaneously from the $\rho_s$ and $VJHK_s$ magnitudes, via two empirical relations of mass-luminosity \citep{2000A&A...364..217D} and diameter-color-luminosity \citep{2004A&A...426..297K}.
As a result, they derived the stellar mass and radius as  $M_s = 0.539^{+0.047}_{-0.043}$ $M_\odot$ and $R_s = 0.568^{+0.037}_{-0.031}$ $R_\odot$, respectively. 
However, the latter relation they adopted was derived from dwarf samples that contained a relatively small number ($<$ 15) of low-mass (K-M) dwarfs.
In addition, the metallicity dependence of this relation was not found in the samples \citep{2004A&A...426..297K}; however, \citet{2012ApJ...757..112B} have recently found, based on  interferometric radius measurements for 33 single K-M dwarfs, that a color-radius relation for low-mass stars clearly depends on metallicity.  Therefore, the relation adopted by D13 could have a systematic offset.
In Figure \ref{fig:MsRs}, we plot a mass-radius relation for low-mass stars provided by \citet[][green shaded region]{2012ApJ...757..112B}, which was derived based on the radius measurements for the 33 K-M dwarfs and the empirical mass-luminosity relation of \citet{1993AJ....106..773H}, which is consistent with that of \citet{2000A&A...364..217D}. we also plot a mass-radius relation for GJ3470 drawn by connecting the two empirical relations that D13 adopted (light-blue shaded region), where $K_s = 7.989$ is used for the mass-luminosity calibration and $H=8.206$ and $(V-H) = 4.124$ are used for the diameter-color-luminosity calibration.
These two relations do not cross each other, implying that some systematic offset could exist, possibly in the diameter-color-luminosity relation of \citet{2004A&A...426..297K}  due to the small number of low-mass star samples and the unconsidered metallicity dependence.

Therefore, unlike D13, we solve $M_s$ and $R_s$ from $\rho_s$ by using the mass-radius relation of \citet[][Equation (4) in their paper]{2012ApJ...757..112B}. 
We note that no metallicity dependence of this relation was detected \citep{2012ApJ...757..112B}. 
In order to properly estimate $M_s$, $R_s$ and their uncertainties, we create probability distributions of these parameters by using a Monte Carlo (MC) technique; that is, the probability distributions are created by repeating a process where $M_s$ and $R_s$ are solved for from a parameter set of $a/R_s$ and three coefficients of the mass-radius relation that are chosen according to their respective probability distributions.
For the probability distribution of $a/R_s$, the posterior distribution created from the MCMC analysis is used, while for those of the three coefficients, Gaussian distributions whose standard deviations are set to their 1-$\sigma$ errors are used.
The resultant median values and the 68.3\% confidence intervals are calculated as $M_s = 0.557^{+0.028}_{-0.020}$ $M_\odot$ and $R_s = 0.526^{+0.023}_{-0.011}$ $R_\odot$.
In Figure \ref{fig:MsRs}, we also plot the derived $M_s$ and $R_s$ values (red circle) as well as those reported by D13 (brown triangle). Our $M_s$ and $R_s$ values are both consistent with those by D13 within 1 $\sigma$. However, this agreement is coincidental, because the derived stellar density is inconsistent between this work and D13 by more than 1 $\sigma$, and the adopted calibration methods for deriving stellar mass and radius are different and also inconsistent between the two. 
Thus, further follow-up observations to confirm the stellar mass and radius would be valuable; photometric transit observations could test the stellar density, and astrometric observations for measuring the trigonometric parallax would provide a new insight into the stellar mass/radius calibration.

Finally, using the estimated $M_s$ and $R_s$, we derive the relevant stellar and planetary parameters. 
For the planetary radius, we adopt the $R_p/R_s$ value in the $J$ band.
This is because not only does the $J$-band light curve have the highest photometric precision among the four light curves, but also, as is discussed in Section \ref{sec:radius_ratio}, the planetary radii in the optical bands could be enlarged compared to $J$ band due to the larger opacity of putative atmospheric haze, and the smaller $J$-band planetary radius could represent a haze-independent planetary radius.
For the planetary mass, we use the following relation assuming a circular orbit: 
$K'$ $\equiv K\sqrt{1-e^2} P^{1/3} = (2\pi G)^{1/3} M_p \sin i / (M_s + M_p)^{2/3}$, where $K$ is the radial-velocity semi-amplitude, and $K' =13.4 \pm 1.2$ m~s$^{-1}$~day$^{1/3}$ is adopted from D13. 
The $M_p$ and $R_p$ values as well as other stellar and planetary parameters  (the stellar surface gravity, $\log g_s$, the semi-major axis, $a$, the planetary density, $\rho_p$, and the planetary surface gravity, $g_p$) are derived by the same MC fashion as was used previously. 
The derived values and uncertainties are listed in Table \ref{tbl:mc}.
The planetary mass, radius, and density are derived as $M_p = 14.1 \pm 1.3$ $M_\oplus$, $R_p = 4.32^{+0.21}_{-0.10}$ $R_\oplus$, and $\rho_p = 0.94 \pm 0.12$ g~cm$^{-1}$, respectively. 
Although the planetary density is $\sim$30\% larger than that by D13 ($0.72^{+0.13}_{-0.12}$ g cm$^{-1}$), it is still well below that of Uranus (1.27 g~cm$^{-1}$) despite their similar masses (14.54 $M_\oplus$ for Uranus). Therefore, we confirm that GJ3470b has a low density, which was first suggested by \citet{2012A&A...546A..27B} and recently established by D13.
We note that the $\sim$30\% difference of the planetary density comes mainly from the $\sim$10\% smaller planetary radius (the $\sim$7\% smaller stellar radius and the $\sim$3\% smaller planet-star radius ratio) compared to D13.
In Table \ref{tbl:mc}, we also list the results derived from the MCMC analysis with $b$ prior for reference.

In order to investigate the impact of the 10\% difference of the planetary radius on the planetary bulk composition, 
we have modeled the internal structure of GJ3470b that is consistent with the planetary mass and radius in the $J$ band derived in this study. To do so, we have assumed that the planet is composed of three layers, namely, a cloud/haze-free hydrogen-rich atmosphere with solar composition, on top of an ice/water mantle, on top of a rocky core (the water/rock mass ratio is set to 3). 
A two-layered structure (hydrogen-rich atmosphere + rocky core) is also examined. 
The equations of state adopted are SCvH EOS for hydrogen/helium \citep{1995ApJS...99..713S}, R-EOS \citep{2009PhRvB..79e4107F} and SESAME EOS \citep{Lyon1992} for water, and \citet{2007ApJ...656..545V} for rock (details are described in K.~Kurosaki et al. in preparation). 
We have found that the mass fraction of the hydrogen-rich atmosphere ranges from 5\% to 9\% in the three-layer models  and from 12\% to 19\% in the two-layer models. 
These values are similar with those derived in D13.
Therefore, the  10\% difference in radius does not alter their conclusion that the planet possesses a hydrogen-rich envelope whose mass accounts for approximately 10\% of the planetary total mass. 
We note that this conclusion may be consistent with recent theoretical prediction: if the atmosphere embedded in a protoplanetary disk grows in mass beyond 10\% of the planetary total mass, then its accretion tends to proceed in a runaway fashion. Thus, nebular-origin atmospheres with intermediate masses, namely, 10\% to several 10\% of planetary mass are rarely detected \citep{2012ApJ...753...66I}.

\subsection{Wavelength Dependence of the Planet-star Radius Ratio} 
\label{sec:radius_ratio}

\begin{figure*}
\begin{center}
\includegraphics[width=13.5cm]{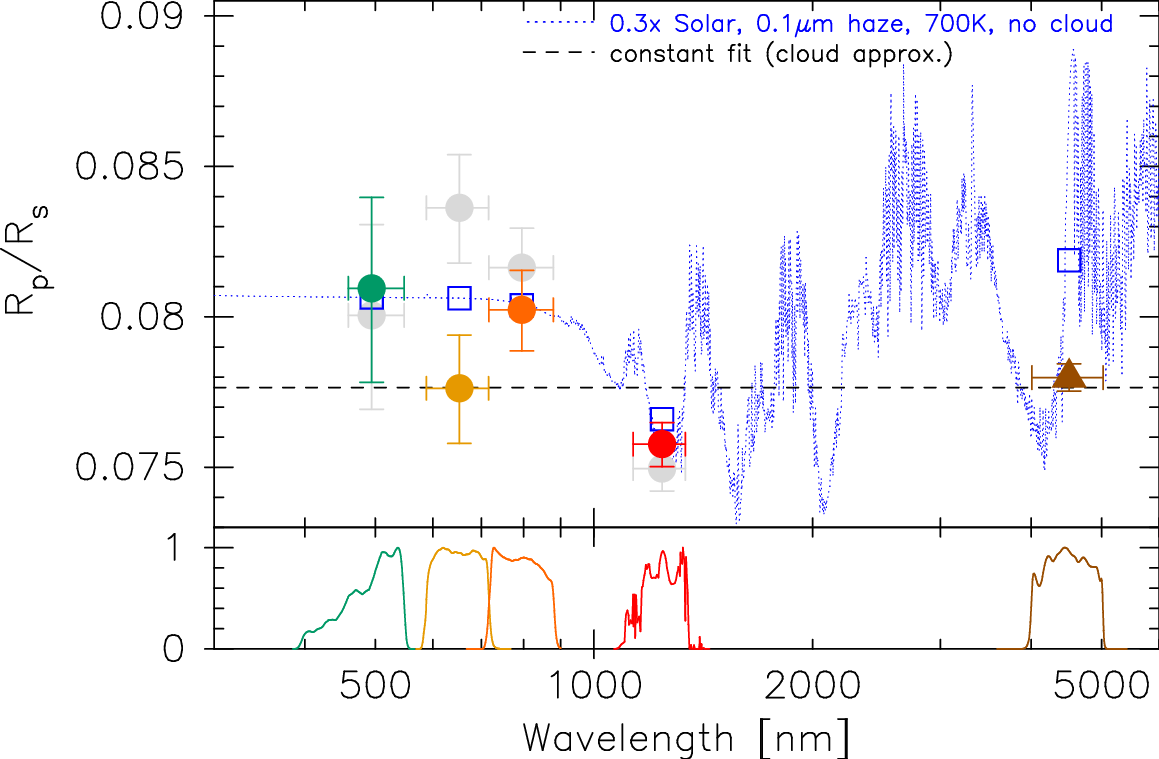}
\caption{Top panel: wavelength dependence of the observed $R_p/R_s$ values for GJ3470b. The data derived from the MCMC analysis in this work with $b$ prior are indicated as filled circles, while that derived by D13 with the $Spitzer$/IRAC 4.5-$\mu$m band is shown as a triangle. The horizontal values and error bars denote the weighted center and the width at half maximum, respectively, of the respective transmission curves shown in the bottom panel: 494$^{+54}_{-35}$nm, 653$^{+63}_{-65}$nm, 796$^{+83}_{-79}$, 1241$^{+95}_{-109}$nm, 4510$^{+512}_{-502}$nm for the $g'$, $R_\mathrm{c}$, $I_\mathrm{c}$, $J$, and 4.5 $\mu$m bands, respectively. The horizontal dashed line indicates the inverse-variance-weighted mean of the five $R_p/R_s$ values (=0.07765), representing an approximated fully-clouded atmospheric spectrum.  The blue dotted line indicates a model spectrum for a 0.3 $\times$ Solar-abundance cloud-free atmosphere with $T=700$ K and containing 0.1-$\mu$m sized tholin particles with a density of 1000 cm$^{-3}$, being drawn by scaling a model spectrum provided by \citet{2012ApJ...756..176H} in order to fit the scale height of GJ3470b.  The blue squares indicate the integrated cloud-free model spectrum over the respective passbands.  The gray circles indicate the impact of possible systematics due to the baseline-model selection; the original value is shifted by the difference of $R_p/R_s$ that would be produced if an alternative baseline model was adopted. Bottom panel: the transmission curves for the respective bands. Those for the $g'$, $R_\mathrm{c}$, and $I_\mathrm{c}$ bands take into account the instrumental transmittance and the CCD's quantum efficiency. 
\label{fig:RpRs}}
\end{center}
\end{figure*}

In this section, we discuss the wavelength dependence of $R_p/R_s$ based on the results of the MCMC analysis with $b$ prior.
In Figure \ref{fig:RpRs}, we plot the $R_p/R_s$ values in the $g'$, $R_\mathrm{c}$, $I_\mathrm{c}$, and $J$ bands, and that in the $Spitzer$'s 4.5-$\mu$m band reported by D13, along with the transmission curves for the respective passbands. 
The $R_p/R_s$ values in the $I_\mathrm{c}$ and 4.5-$\mu$m bands are larger than that in the $J$ band by $5.8\% \pm 2.0\%$ and $2.9\% \pm 1.1\%$, respectively. 
Those in the $g'$ and $R_\mathrm{c}$ bands are also larger than that in the $J$ band, although the uncertainties in these bands are relatively large.

As discussed in Section \ref{sec:reduction_isle} and \ref{sec:reduction_mitsume}, the choice of the baseline model could cause a systematic offset in $R_p/R_s$. However, the alternative choices of the baseline models for the $R_\mathrm{c}$-, $I_\mathrm{c}$-, and $J$-band light curves expand the discrepancies, as indicated by gray circles in Figure \ref{fig:RpRs}, meaning that the systematics due to the baseline selections cannot explain the observed discrepancies of $R_p/R_s$.

\begin{figure}
\includegraphics[width=8.5cm]{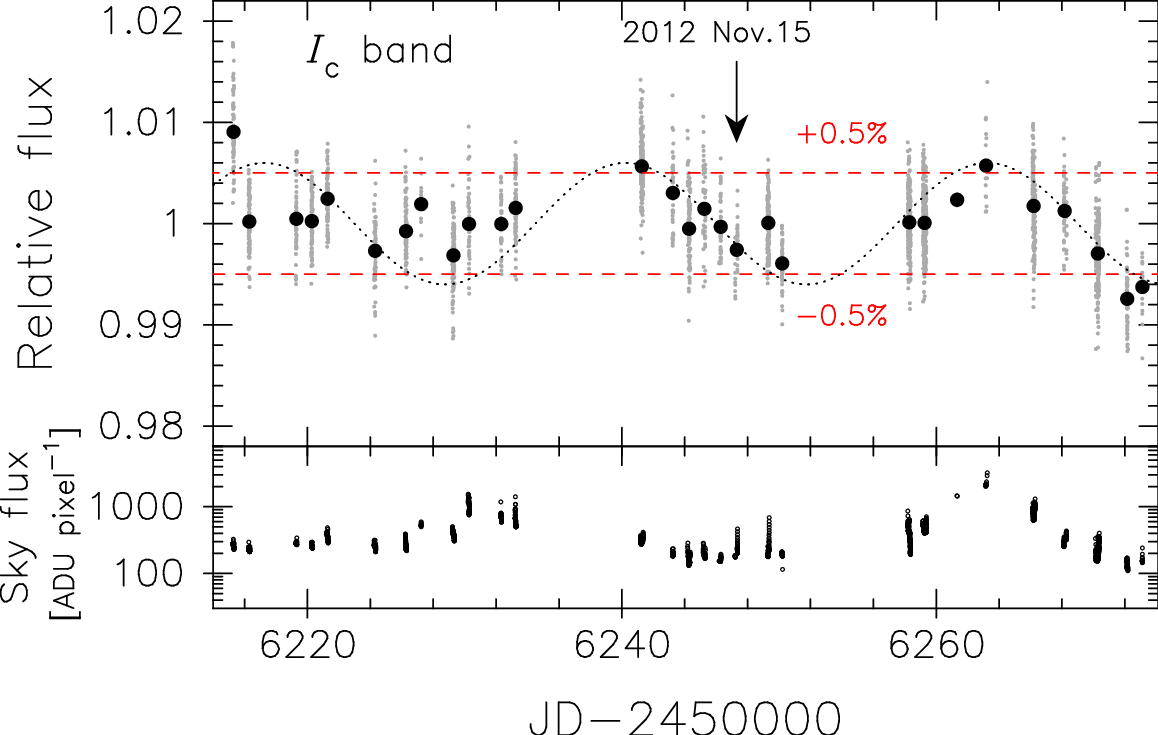}
\caption{Top panel: a two-month-long $I_\mathrm{c}$-band light curve for GJ3470 obtained by the 50-cm MITSuME telescope. Transit parts are eliminated in this light curve. Individual 60 s exposure data and nightly-averaged data are indicated as gray and black points, respectively. The rms of the black points is 0.33\%, while peak-to-valley variations is $\sim$1\%. The $\pm$0.5\% variation level is shown as dashed line. The transit-observed date is indicated by an arrow. The dotted auxiliary line is for a 23-days periodicity, part of which could be caused by sky-flux variations. Bottom panel: time variations of the sky flux. A correlation between the light curve and the sky flux can be seen after JD-2,450,000 $\sim$ 6258.\label{fig:longlc}}
\end{figure}

Unocculted starspots can also vary apparent $R_p/R_s$ with time, due to the variations of the apparent luminous area of the star according to the stellar rotation and appearing/vanishing of the starspots \citep[e.g.][]{2008MNRAS.385..109P}. 
Since the observational epochs are different with $\sim$5 months apart, there is a possibility that the observed $R_p/R_s$ difference between the $J$ and 4.5-$\mu$m bands is  caused by the starspot effect.
When a spotted host star induces the variability of $r \equiv R_p/R_s$ by $\Delta r$ due to the unocculted star-sopt effect, the host star has to show an intrinsic flux variability of at least $(1+\Delta r/r)^2 -1$.
Therefore, if the $2.9\% \pm 1.1\%$ difference of $R_p/R_s$ between the $J$ and 4.5-$\mu$m bands were caused by unocculted starspots, the host star would have to show at least $5.9\% \pm 2.3\%$ intrinsic variability.
In order to estimate the maximum intrinsic variability of the host star, we conducted a photometric monitoring of GJ3470 for about two months by using the 50-cm MITSuME telescope. The resultant $I_\mathrm{c}$-band light curve is shown in Figure \ref{fig:longlc}, along with the sky-flux variations. We find that the rms of the nightly averaged fluxes is 0.33\% and the peak-to-valley variability is $\sim$1\%, indicating that GJ3470 is not a very active star and starspot variations cannot account for the observed $R_p/R_s$ difference between the $J$ and 4.5-$\mu$m bands, unless the stellar variability is unusually large only in infrared wavelength.
In addition, the difference of $R_p/R_s$ between the $I_\mathrm{c}$  and $J$ bands cannot intrinsically be explained by the spot-rotation effect because these data were obtained simultaneously.
We note that although the two data sets were obtained simultaneously,   in principle, there is still a possibility that a chromatic difference of spot brightness could vary the $R_p/R_s$ values in the two bands by a small amount \citep{2008MNRAS.385..109P}. However, this effect is tiny even for an active host star ($< 10^{-4}$ in $R_p/R_s$), and is negligible in this case because the host star is not likely to be an active star.   
We also note that we find a possible $\sim$23-day periodic variation in the monitoring light curve (dotted line in Figure \ref{fig:longlc}). This periodicity could potentially be due to starspot rotation according to the stellar rotation. However, we also find that a part of the flux variations (JD $\gtrsim$ 2,456,258) could be correlated with the sky-flux variations, and therefore the $\sim$23-day periodicity is suspicious and further monitoring is needed to confirm it (more will be investigated in a future paper).

A more likely scenario for the $R_p/R_s$ variations would be that the planetary atmospheric opacity varies with wavelength due to absorption and/or scattering by atmospheric molecules.
The relative change in transit depth due to molecular absorption as a function of wavelength can be approximated by $\Delta D (\lambda)  \sim 2 n_\mathrm{H} (\lambda) H R_p/R_s^2$,
where $n_\mathrm{H}(\lambda)$ is the scale factor depending on the molecular opacity, which can be $\sim$10 for strong absorbers \citep{2001ApJ...553.1006B},   and $H$ is the atmospheric scale height given by $H \equiv kT/\mu g_p$, where $k$ is Boltzmann's constant, $T$ is the atmospheric temperature, and $\mu$ is the mean molecular weight. For GJ3470b, H$\sim$250--400 km when assuming $T=500$--650K and a solar-composition atmosphere ($\mu = 2.36$ atomic mass). Therefore, $\Delta D (\lambda)$ can be up to $\sim$0.16\%, which corresponds to a relative change in $R_p/R_s$ of $\sim$$13\%$, indicating that the observed differences of $R_p/R_s$ can reasonably be explained by molecular absorptions.

Indeed, there is a strong absorption band of CO around 4.7 $\mu$m, and the CO-absorption feature in $R_p/R_s$ becomes to be prominent  for a solar-abundance atmosphere with $T \gtrsim$ 700 K \citep{2012ApJ...756..176H}.
The planetary-equilibrium temperature is estimated as $(1-A)^{1/4} (683 \pm 27)$ K (D13), where $A$ is bond albedo. With this temperature, although this is lower than 700 K, it could be possible that the CO feature could be seen at some level. Therefore, the observational result that $R_p/R_s$  in the 4.5 $\mu$m band is larger than that in the $J$ band is qualitatively consistent with a CO-containing atmosphere. 
On the one hand, the larger $R_p/R_s$ values in the optical ($I_\mathrm{c}$, $R_\mathrm{c}$, and $g'$) bands compared to that in the $J$ band could be explained by Rayleigh scattering due to molecular hydrogen or small-sized ($\lesssim$0.1 $\mu$m) haze in the atmosphere, as is the case for HD~209458b \citep{2008A&A...485..865L} and HD~189733b \citep{2008MNRAS.385..109P}. The transition wavelength below which the spectrum begins to rise due to Rayleigh scattering is around $1\mu$m for $\sim$0.1-$\mu$m sized haze particles \citep[e.g.][]{2012ApJ...756..176H}, and therefore the observed $R_p/R_s$ difference between the $I_\mathrm{c}$ and $J$ bands could be explained well by such a hazy atmospheric model.
In Figure \ref{fig:RpRs}, we also plot a model spectrum (blue dotted line), which is selected among those for the atmosphere of a 10-$M_\oplus$ planet provided by \citet{2012ApJ...756..176H} so as to qualitatively fit the data, and is scaled in order to fit the scale height of GJ3470b. The selected model assumes a 0.3 $\times$ solar-abundance atmosphere with $T=700$K containing 0.1-$\mu$m tholin particles with a density of 1000 cm$^{-3}$.
We note that this model does not explain the data point at 4.5-$\mu$m band well, however, in reality the CO-absorption feature at $\sim$4.7 $\mu$m 
is probably weaker than the model because the planet is a bit cooler than the temperature that the model assumes.
A quantitative discussion and fine-tuning of this model are beyond the scope of this paper, and to do so, further observational data are required.

In contrast, if the atmosphere of GJ3470b were covered by thick clouds at high altitudes, then a flat spectrum would be observed over optical to infrared wavelengths \citep[e.g.][]{2012ApJ...747...35B}. 
A constant fit to the observed five $R_p/R_s$ values yields a $\chi^2$ value of 11.6. The statistical probability that $\chi^2$ exceeds this value assuming Gaussian errors is 2.1\%, meaning that a fully-clouded atmospheric model still cannot be ruled out by the current data. Therefore, further observations are needed to confirm the possible $R_p/R_s$ variations over optical to infrared wavelengths.

If the $R_p/R_s$ variations are confirmed, this fact will indicate that the planet would have no thick clouds in its atmosphere.
 This property would offer a wealth of opportunity to probe certain molecular features in the atmosphere of GJ3470b through high-precision transmission spectroscopic observations, without being prevented by thick clouds. 
Specifically, molecular features such as H$_2$O, CH$_4$, and CO could easily be detected in infrared-wavelengths range, assuming a solar-like atmospheric composition. The detection of H$_2$O would imply that there would be a possibility that this planet would have originally formed beyond the snow line and later migrated to the current position. Measuring the C/O ratio from these molecular features would also be interesting not only in the planetary-formation point of view, but also in the chemical-characterization point of view for planetary atmospheres \citep{2012ApJ...758...36M}.

We should note that this is the first report on transit observations by using the 188-cm telescope/ISLE instrument and the 50-cm MITSuME telescope. Our observations for GJ3470b demonstrate that these telescopes/instruments are useful for studying transiting planets, especially for probing planetary atmospheres by simultaneous observations through the optical to near-infrared wavelengths. Nevertheless, the capability of the 50-cm MITSuME telescope is limited due to the limited aperture size. It would be a great capability if 2-4 m class telescopes like the 188-cm telescope were equipped with a transit-dedicated camera that could take  multi-color images through optical to infrared simultaneously.
The GROND instrument, mounted on the MPG/ESO 2.2-m telescope in Chile \citep{2008PASP..120..405G}, is a pioneer for such an ambitious instrument, which can obtain seven images through the $g'$ to $K$ bands simultaneously for the southern hemisphere.  
Recently, simultaneous transit observations by using the GROND instrument have been reported \citep{2013arXiv1301.3005M,2012A&A...538A..46D}, demonstrating the usefulness of such a multi-color imager for transit observations. 
However, this instrument had originally been developed for catching gamma-ray-burst afterglows and was not designed for high-precision transit observations \citep{2013arXiv1301.3005M}.
Future developments of such instruments, but more specifically designed for transit observations, for 2-4 m class telescopes in the northern hemisphere will provide fruitful results on the exoplanetary atmospheric studies.

\section{SUMMARY}
\label{sec:summary}

We present optical ($g'$, $R_\mathrm{c}$, and $I_\mathrm{c}$) to near-infrared ($J$) simultaneous photometric observations for a primary transit of the hot Neptune GJ3470b, by using the 188-cm telescope/ISLE instrument and the 50-cm MITSuME telescope, both at Okayama Astrophysical Observatory.
 We found that the planetary density is broadly consistent with that reported by \citet{2013arXiv1301.6555D} who measured it based on the $Spitzer$/IRAC 4.5-$\mu$m photometry, confirming its low density. 
 Although the derived planetary radius is about 10\% smaller than that reported by D13,
this difference does not alter their conclusion that the planet possesses a hydrogen-rich envelope whose  mass is approximately 10\% of the planetary total mass. 
On the other hand, we have found that the planet-to-star radius ratio ($R_p/R_s$) in the $J$ band is smaller than that in the $I_\mathrm{c}$ and 4.5-$\mu$m bands by $5.8\% \pm 2.0\%$ and $2.9\% \pm 1.1\%$, respectively.
These discrepancies cannot be explained by systematic effects due to baseline corrections.
In addition, we have found from a two-month-long flux monitoring of GJ3470 that the intrinsic peak-to-valley stellar variability in the $I_\mathrm{c}$ band is mealy $\sim$1\%, indicating that the unocculted starspot effect is unlikely to account for the observed $R_p/R_s$ difference between the $J$ and 4.5-$\mu$m bands.
Instead, Rayleigh scattering due to molecular hydrogen or small-sized ($\lesssim$0.1 $\mu$m) haze in the atmosphere and molecular absorptions such as CO could reasonably explain the observed $R_p/R_s$ variations.
Although the significance is low, if these $R_p/R_s$ variations are confirmed, then this fact would suggest that GJ3470b would not have a thick cloud layer in the atmosphere, offering a wealth of opportunity for future transmission-spectroscopic observations to probe many molecular features in the atmosphere of the hot Neptune.

We note that this is the first report on transit observations using the 188-cm telescope/ISLE instrument and the 50-cm MITSuME telescope. Our observations demonstrate that optical-to-near-infrared simultaneous observations using such as these are very useful for planetary atmospheric studies. Future developments of multi-color imagers for 2-4 m class telescopes in the northern hemisphere would provide much more fruitful results for this field.

\acknowledgments
We thank Y. Hori for valuable discussions.
We acknowledge a support by NINS Program
for Cross-Disciplinary Study. N.N. is supported by
NAOJ Fellowship and by the JSPS Grant-in-Aid for
Research Activity Start-up No. 23840046.
Y.H.T and T.H are supported by JSPS Fellowships for Research (DC1: 23-271 and 22-5935, respectively).
We are grateful to the anonymous referee for an in-depth review and thoughtful comments, including many corrections that significantly improved the manuscript.





\end{document}